\documentclass{article}

\usepackage{amssymb,amsmath,times}

\DeclareMathAlphabet{\altmathcal}{OMS}{cmsy}{m}{n}

\usepackage{arydshln}
\usepackage[pdftex]{graphicx}
\usepackage{hyperref}

\newtheorem{theorem}{Theorem}
\newtheorem{lemma}[theorem]{Lemma}

\newcommand\vartextvisiblespace[1][.9em]{%
  \makebox[#1]{%
    \kern.07em
    \vrule height.9ex
    \hrulefill
    \vrule height.9ex
    \kern.07em
  }
}

\newcommand{\xvdash}[1]{%
  \vdash^{\mkern-10mu\scriptstyle\rule[-.9ex]{0pt}{0pt}#1}%
}

\begin{document}

\title{Self-referential basis of undecidable dynamics: from The Liar Paradox and The Halting Problem to The Edge of Chaos}
\author{
Mikhail Prokopenko$^{1}$, Michael Harr\'e$^{1}$, Joseph Lizier$^{1}$,\\
Fabio Boschetti$^{2}$, Pavlos Peppas$^{3,4}$, Stuart Kauffman$^{5}$\\ 
\\
$^{1}$Centre for Complex Systems, Faculty of Engineering and IT\\ 
			The University of Sydney, NSW 2006, Australia\\
$^{2}$CSIRO Oceans and Atmosphere, Floreat, WA 6014, Australia\\
$^{3}$Department of Business Administration, University of Patras, Patras 265 00, Greece\\
$^{4}$University of Pennsylvania, Philadelphia, PA 19104, USA\\
$^{5}$University of Pennsylvania, USA\\
        mikhail.prokopenko@sydney.edu.au
        }

\date{}

\maketitle

\begin{abstract}
In this paper we explore several fundamental relations between formal systems, algorithms, and dynamical systems, focussing on the roles of undecidability, universality, diagonalization, and self-reference in each of these computational frameworks.  Some of these interconnections are well-known, while some are clarified in this study as a result of a fine-grained comparison between recursive formal systems, Turing machines, and Cellular Automata (CAs). In particular, we elaborate on the diagonalization argument applied to distributed computation carried out by CAs, illustrating the key elements of G\"{o}del's proof for CAs. The comparative analysis emphasizes three factors which underlie the capacity to generate undecidable dynamics within the examined computational frameworks: (i) the program-data duality; (ii) the potential to access an infinite computational medium; and (iii) the ability to implement negation. The considered adaptations of G\"{o}del's proof distinguish between computational universality and undecidability, and show how the diagonalization argument exploits, on several levels, the self-referential basis of undecidability. 
\end{abstract}

%\begin{keyword}
%%% keywords here, in the form: keyword \sep keyword
%
%self-reference \sep diagonalization \sep undecidability \sep incomputability \sep program-data duality \sep complexity
%
%%% PACS codes here, in the form: \PACS code \sep code
%
%%% MSC codes here, in the form: \MSC code \sep code
%%% or \MSC[2008] code \sep code (2000 is the default)
%
%\end{keyword}

%\end{frontmatter}

\section{Introduction}

It is well-known that there are deep connections between dynamical systems, algorithms, and formal systems. These connections relate the Edge of Chaos phenomena observed in dynamical systems, to the Halting problem recognized in computability theory, as well as to G\"{o}del's Incompleteness Theorems established within the framework of formal systems.   Casti, for example, has explored interconnections between dynamical systems, G\"{o}delian formal logic systems, Turing machines, as well as Chaitin's complexity results, arguing that 
\begin{quote}
``the theorems of a formal system, the output of a UTM [Universal Turing Machine], and the attractor set of a dynamical process (e.g., a 1-dimensional cellular automaton) are completely equivalent; given one, it can be faithfully translated into either of the others.'' \cite{Cas91}.
\end{quote}
A similar triangle of equivalences between Physics (dynamical systems), Mathematics (formal systems) and Computation (algorithms) is discussed by Ilachinski in the context of the Anthropic Principle :
\begin{quote}
``Just as G\"{o}del's theorem makes use of logical self-reference to prove the existence of unprovable truths within a mathematical system, and Turing's theorem makes use of algorithmic self-reference to show that a computer cannot fully encompass, or understand, itself, the anthropic principle limits the perceived structure of the universe by the fact that the universe is effectively perceiving itself.'' \cite{Ilachinski2001}.
\end{quote}
These arguments bring forward several key concepts which underlie the analogies --- \emph{undecidability}, \emph{universality} and \emph{self-reference} --- and implicate them in the notions of  \emph{chaos} and \emph{complexity}.  

An undecidable problem is typically defined in computability theory as a decision problem for which it can be shown that a correct yes-or-no answer cannot always be produced by an algorithm. One of the most well-known examples of undecidable problems is the Halting problem: given a description of an arbitrary program (e.g., a Turing machine) and an input, it is impossible to construct an algorithm which would determine whether the program will eventually halt or continue to run forever. In the context of formal logic systems, an undecidable statement is a statement expressible in the system's language  which can neither be proved nor disproved within the very same system. The phenomenon of undecidability is present in dynamical systems as well, and needs to be distinguished from deterministic chaos:
\begin{quote}
``For a dynamical system to be chaotic means that it exponentially amplifies ignorance of its initial condition; for it to be undecidable means that essential aspects of its long-term behaviour --- such as whether a trajectory ever enters a certain region --- though determined, are unpredictable even from total knowledge of the initial condition.''  \cite{Bennett1990},
\end{quote}
where the behavior is meant to be unpredictable without full simulation.
While describing an example of undecidable dynamics of a physical particle-motion system with mirrors, Moore has also distinguished between ``sensitive dependence'' and ``algorithmic complexity'': in the former case the chaotic dynamics are unpredictable due to imperfect knowledge of initial conditions, while in the latter case (undecidability), ``even if the initial conditions are known exactly, virtually any question about their long-term dynamics is undecidable'' \cite{Moore1990,Moore1991}. 
A very well-studied type of discrete dynamical systems where the classes of ordered, chaotic and complex (``Edge of Chaos'') dynamics have been identified and characterized is Cellular Automata (CAs), although the ability to quantitatively separate such classes is often questioned \cite{Durand,Kari,Sutner}.  Computationally, CAs can be seen as information-processing systems  carrying out a computation on data represented by an initial configuration  \cite{Wolfram1984}. Being a computational device, a CA may also be analyzed in terms of undecidable dynamics (although one must carefully specify what questions are put to a test), and such an analysis invariably involves the concept of computational  universality \cite{wolfram1985,Wolfram2002}.  

As pointed out by Bennett \cite{Bennett1990}, ``a discrete or continuous dynamical system is called computationally universal if it can be programmed through its initial conditions to perform any digital computation'', and moreover, ``universality and undecidability are closely related: roughly speaking, if a universal computer could see into the future well enough to solve its own halting problem, it could be programmed to contradict itself, halting only if it foresaw that it would fail to halt.'' This succinct phrase emphasizes that undecidability is a consequence of universality, and reaches to the core of the self-referential argument utilized in demonstrating undecidability within various computational frameworks. 

This brings us to one of the central objectives of this work --- elaborating on the role played by self-reference in distributed computation carried out by CAs.

The Liar's Paradox which has captured the imagination of philosophers and logicians for thousands of years is a self-referential statement the truth or falsity  of which cannot be assigned without a contradiction:  for example, the paradox can be presented as a statement of a person declaring that ``everything I say is a lie'', or  more formally as ``this statement is unprovable''.  It has achieved prominence in modern philosophical logic largely due to the motivation it provided to various proofs of incompleteness, undecidability, and incomputability. 
A fundamental aspect of this paradox, and the works which incorporated its main idea, is self-reference: the way the statement refers to its own validity.   
As we shall see, there is a close but subtle difference between the concept of self-reference and the diagonalization argument (dating back to Cantor's diagonal argument), both of which play important roles in formal systems, algorithms, and dynamical systems. 

Despite the early realization of  fundamental interconnections between formal systems, algorithms (Turing machines), and dynamical systems, the precise set of detailed analogies remains elusive, leading sometimes to inaccurate parallels.
For instance, Casti offers a ``logical route'' to chaos, claiming that ``there is a direct chain of connection linking the existence of strange attractors, Chaitin's results on algorithmic complexity, and G\"{o}del's Incompleteness Theorem''  \cite{Cas91}.  As he points out, Cellular Automata theorists, while distinguishing between ``strange attractors'' and ``quasiperiodic orbits'',  ``lump both types into the same category of ``strange attractor'' when trying to make contact with the traditional dynamical systems literature'' \cite{Cas91}. Obviously, the analysis presented by Casti has since been further illuminated  by studies of  class IV CAs (``quasiperiodic orbits''), highlighting the differences between their  complex dynamics at the edge of chaos from  class III CAs (``strange attractors'')  \cite{lang90,crutch94,wue99,hord01,Wolfram2002,sha06,liz08a,liz12a}.

It has been long-conjectured that ``complex'' systems evolve to the ``edge of chaos'', that is, their dynamical behavior is neither \emph{ordered}, i.e., globally attracting a fixed point or a limit cycle, nor \emph{chaotic}, i.e., sensitive to imperfectly known initial conditions \cite{pack88,Crutchfield1988,lang90,Wolfram2002}. These broad claims have been questioned, and indeed it has been demonstrated that computational tasks can certainly be achieved away from the edge of chaos \cite{mitch94b}. A more appropriate interpretation, without claims appealing to evolution, may be that (i) while all classes of systems undertake intrinsic computation (and indeed the most appropriate type of system for handling particular computational tasks may be distant from the edge of chaos \cite{mitch94b}), (ii) there is evidence that the edge of chaos offers computational advantages (e.g. blending information storage and transfer capabilities) that are advantageous for a priori unknown or indeed general purpose computational tasks \cite{liz08c,liz11b,boed12a}. Ilachinski also directly mapped (a) halting computation of CAs to  class I (``frozen'' dynamics, i.e., fixed-points) and  class II (periodic dynamics, i.e., limit cycles); (b) non-halting computation to class III (chaotic dynamics, i.e., ``strange attractors''), and (c) undecidable computation to class IV (``arbitrarily long transients'') \cite{Ilachinski2001}. Nevertheless, it has also been argued that some chaotic systems may also be universal, and hence, not decidable, contradicting the thesis that universal computation can only happen at the ``edge of chaos'', while acknowledging that the existence of a chaotic universal CA has not yet been demonstrated and remains an open question \cite{Siegelmann1999,delvenne2006decidability,Zenil2013}.

However, the difficulty in identifying what kind of  CA dynamics corresponds to the undecidability appears  not only due to the lack of a standard classification, but also due to different computational structures employed by CAs and say, Turing machines. In particular, one needs to take special care in drawing parallels between a CA running on an initial configuration, on the one hand, and a formal system inferring theorems from a set of axioms, on the other hand. Indeed, undecidable statements of a formal system which may be more akin to  ``quasiperiodic orbits'' (class IV CAs) rather than ``strange attractors'' (class III CAs), might be so only with respect to a given initial configuration. Furthermore, in order to relate the attractors of CAs dynamics to the outcomes of Turing machines, or to  the theorems derived by formal systems, a consideration must be given to carefully setting up a termination condition for CAs.

While the program-data duality allows us to freely move elements of a computational system between the \emph{program} (a CA's rule-table, a Turing machine's transition function, or a formal system's rules of inference) and the \emph{data} (a CA's initial configuration, a Turing machine's input tape, a formal system's axioms), the type of the eventual dynamics and hence, a possible classification, depends on both components.
Thus, a classification scheme which, in principle, aims to classify a program running on \emph{all} inputs, cannot distinguish between the types corresponding to halting, non-halting and undecidable decision problems which are specifically defined for a system with both program and data. The classification problem itself has been shown to be undecidable for a broad range of cases \cite{Durand,Kari,Sutner-chapter,Sutner}.

Finally, while the key role played by the self-reference in proofs of undecidability in various computational frameworks is beyond doubt, its precise use in dynamical systems, and CAs specifically, has not been demonstrated explicitly. As discussed by  \cite{Hyotyniemi}, in a dynamical system, the Liar's paradox may take the following form: ``the system is not stable if and only if it can be shown to be stable''. This analogy is not a perfect equivalence, as it simply entails that  there is no method for determining the stability of such a system \cite{Hyotyniemi}. However, rather than pointing out that a dynamical system is computationally equivalent to an algorithm and then  restating the paradox  in the language of dynamical systems, it could be more elucidating to  constructively demonstrate how and where self-reference is implicated in the structure and dynamics of a CA.  

Such an undertaking is the main focus of our study: without engaging in a philosophic debate on the nature of the self-reference (which continues to be vigorously discussed in modern philosophical logic), we shall attempt to essentially reconstruct main elements of G\"{o}del's proof for Cellular Automata.  In doing so, we shall find more precise and fine-grained parallels between the key elements of three computational frameworks (formal systems, Turing machines, Cellular Automata), some of which have been pointed out previously  \cite{Cas91,Ali1999,Ilachinski2001}, while some have become apparent as a result of the direct comparison between the respective adaptations of G\"{o}del's proof. These adaptations, we hope, can serve the second purpose of this study, aiming to make G\"{o}del's proof and the related concepts of self-reference, diagonalization, universality and undecidability more accessible to the cross-disciplinary field of Complex Systems.

\section{Methods}

\subsection{Formal Systems and The Liar Paradox}
\label{FS}
\vspace*{2mm}

\subsubsection{Technical preliminaries}

We shall briefly define formal systems in order to formulate the Liar's Paradox and establish the connections to self-reference and diagonalization. In doing so, we shall begin with original definitions of mathematical and elementary formal systems by Smullyan \cite{Smullyan1961}, which utilise the concept of well-formed formulas built from some symbols.  Then we extend the definition of a formal system with a grammar component which specifies how well-formed formulas are constructed in general.
 
Following Smullyan \cite{Smullyan1961}, we can define a mathematical system with at least three  items \[\altmathcal{F} = \langle \altmathcal{A}_{\altmathcal{F}}, \altmathcal{X}_{\altmathcal{F}}, \altmathcal{R}_{\altmathcal{F}} \rangle \]
where
\begin{enumerate}
\item $\altmathcal{A}_{\altmathcal{F}}$ is an alphabet, i.e., an ordered finite set of symbols, so that $\altmathcal{A}^*_{\altmathcal{F}}$ is the set of  words (strings) that can be formed as finite linear sequences of symbols from $\altmathcal{A}_{\altmathcal{F}}$ (i.e., $\altmathcal{A}^*_{\altmathcal{F}}$ is formed by the Kleene operator applied to $\altmathcal{A}_{\altmathcal{F}}$);
\item $\altmathcal{X}_{\altmathcal{F}} \subseteq \altmathcal{A}^*_{\altmathcal{F}}$ is a specific set of axioms;
\item $\altmathcal{R}_{\altmathcal{F}}$ is a finite set  of relations in $\altmathcal{A}^*_{\altmathcal{F}}$ called rules of inference.
\end{enumerate} 
\emph{Axioms} serve as  premises for further inferences, by the \emph{inference rules}, which can be stated in a generic form: 
\[ \mathrm{zero \ or \ more \ premises}   \Rightarrow \mathrm{conclusion} \]
For example, the \emph{modus ponens} rule of propositional logic $a, a \rightarrow b \Rightarrow b$, infers the conclusion $b$ whenever $a$ and $a \rightarrow b$ have been obtained (either as given axioms, or as previous inferences). Axioms and inference rules are used to derive (i.e., prove) theorems of the system.

Typically, an expression $W$ is said to be derivable or formally provable in $\altmathcal{F}$ if and only if there is
a finite sequence of expressions $W_1, \ldots, W_n$ in which  $W \equiv W_n$ and every $W_i$ is either an axiom or results from the application of an inference rule to earlier expressions in the sequence \cite{Buldt2016,Rapaport2005}. We follow the standard notation $\altmathcal{F} \vdash W$ expressing that $W$ is derivable in the formal system $\altmathcal{F}$, in other words that there is a proof of $W$ in $\altmathcal{F}$, i.e., $W$ is a theorem of $\altmathcal{F}$.
However, in order to call $W$ a theorem, one still needs to either apply some external criterion distinguishing $W$ from intermediate derivations in advance, as a target expression, or recognize its standing as having a special salience at the meta-level, capturing it as a theorem (current developments are not able to formally distinguish such salience).

In forming the set of words $\altmathcal{A}^*_{\altmathcal{F}}$ we did not need to follow any additional syntactic constraints, but one may choose to focus only on \emph{well-formed formulas} (abbreviated as wff's), constructed from the alphabet $\altmathcal{A}_{\altmathcal{F}}$ following some grammar.  
The formalization of a grammar $\altmathcal{G}_{\altmathcal{F}} = \langle \altmathcal{A}_{\altmathcal{F}}, \altmathcal{N}_{\altmathcal{F}}, \altmathcal{P}_{\altmathcal{F}}, \altmathcal{S}_{\altmathcal{F}} \rangle$ consists of the following components \cite{Chomsky1956}:
\begin{enumerate}
\item a finite set $\altmathcal{A}_{\altmathcal{F}}$ of terminal symbols;
\item a finite set $\altmathcal{N}_{\altmathcal{F}}$ of nonterminal symbols, that is disjoint with $\altmathcal{A}^*_{\altmathcal{F}}$, i.e., the strings formed from $\altmathcal{A}_{\altmathcal{F}}$;
\item a finite set $\altmathcal{P}_{\altmathcal{F}}$ of production rules of the form $(\altmathcal{A}_{\altmathcal{F}} \cup \altmathcal{N}_{\altmathcal{F}})^{*}\altmathcal{N}_{\altmathcal{F}}(\altmathcal{A}_{\altmathcal{F}} \cup \altmathcal{N}_{\altmathcal{F}})^{*} \rightarrow (\altmathcal{A}_{\altmathcal{F}} \cup \altmathcal{N}_{\altmathcal{F}})^{*}$,
so that each production rule maps from one string of symbols to another, with the ``head'' string containing an arbitrary number of symbols provided at least one of them is a nonterminal;
\item the start symbol $\altmathcal{S}_{\altmathcal{F}} \in \altmathcal{N}_{\altmathcal{F}}$.
\end{enumerate}
The terminal symbols may appear in the output of the production rules but cannot be replaced using the production rules, while nonterminal symbols can be replaced.  For example, the grammar $\altmathcal{G}_{\altmathcal{F}}$ with  
$\altmathcal{N}_{\altmathcal{F}}=\left\{\altmathcal{S}_{\altmathcal{F}}\right\}$,
 ${\altmathcal{A}}_{\altmathcal{F}} =\left\{a,b\right\}$,  and $\altmathcal{P}_{\altmathcal{F}}$ 
with two production rules $\altmathcal{S}_{\altmathcal{F}} \rightarrow a\altmathcal{S}_{\altmathcal{F}}b$ and $\altmathcal{S}_{\altmathcal{F}}\rightarrow ba$, generates wff's $a^nbab^n$, for $n \ge 0$, e.g., $ba$, $abab$, $aababb$, and so on, by applying the first rule $n$ times, followed by one application of the second rule.

%As pointed out by Smullyan \cite{Smullyan1961}, a formal mathematical system $\altmathcal{F}$  can be represented by a predicate $P$ in some elementary formal system $\altmathcal{E}$ which (instead of a grammar) uses  predicates and free variables in constructing its wff's. The representability of a set $x_1, \ldots, x_n$ of free variables by a predicate $P$ of degree $n$ means that  $P(x_1, \ldots, x_n)$ is provable in $\altmathcal{E}$, and so the set of theorems of $\altmathcal{F}$ can be represented by some predicate $P$ in $\altmathcal{E}$.
%In this case, the question as to whether or not a formula $W \in A^*_{\altmathcal{F}}$ is a theorem of $\altmathcal{F}$ is equivalent to the question of whether or not $P(W)$  is a theorem of $\altmathcal{E}$.

Following more recent treatments of formal systems, one may explicitly include components of a grammar  $G$  in the definition \[\altmathcal{F} = \langle \altmathcal{A}_{\altmathcal{F}}, \altmathcal{N}_{\altmathcal{F}}, \altmathcal{P}_{\altmathcal{F}}, \altmathcal{X}_{\altmathcal{F}}, \altmathcal{R}_{\altmathcal{F}} \rangle \]
where
\begin{enumerate}
\item $\altmathcal{A}_{\altmathcal{F}}$ is an alphabet, i.e., an ordered finite set of symbols;
\item  $\altmathcal{N}_{\altmathcal{F}}$ is a finite set of nonterminal symbols, including the start symbol $\altmathcal{S}_\altmathcal{F} \in \altmathcal{N}_{\altmathcal{F}}$, that is disjoint with $\altmathcal{A}_{\altmathcal{F}}^*$;
\item  $\altmathcal{P}_{\altmathcal{F}}$ is a finite set of production rules of the form $(\altmathcal{A}_{\altmathcal{F}} \cup \altmathcal{N}_{\altmathcal{F}})^{*}\altmathcal{N}_{\altmathcal{F}}(\altmathcal{A}_{\altmathcal{F}} \cup \altmathcal{N}_{\altmathcal{F}})^{*} \rightarrow (\altmathcal{A}_{\altmathcal{F}} \cup \altmathcal{N}_{\altmathcal{F}})^{*}$;
\item $\altmathcal{X}_{\altmathcal{F}}$ is a specific set of axioms, each of which must be a wff;
\item $\altmathcal{R}_{\altmathcal{F}}$ is a finite set  of relations in the set of wff's, called rules of inference.
\end{enumerate}
	
That is, while the production rules in $\altmathcal{P}_{\altmathcal{F}}$ are used to produce wff's, the rules of inference in $\altmathcal{R}_{\altmathcal{F}}$ are required to derive theorems.  
We would like to point out that if we consider all words (strings) in $\altmathcal{A}_{\altmathcal{F}}^*$ as wff's, then the grammar would not be constraining the space of possible inferences. In a special case that a formal system contains negation, a system is called consistent if there is no wff $W$ such that both $W$ and $\neg W$ can be proved.

It is usually required that there is a decision procedure (utilizing $\altmathcal{P}_{\altmathcal{F}}$) for deciding whether a formula is well-formed or not. In other words, it is generally assumed that the production rules are decidable: there is an algorithm such that, given an
arbitrary string $x$, it can decide whether $x$ is a wff or not.  Inference rules need also be decidable in the following sense: for
each inference rule $R \in \altmathcal{R}_{\altmathcal{F}}$, there needs to be an algorithm such that, given a set of wff $x_1, \ldots, x_n$ and a wff $y$, the algorithm can
decide if $R$ can be applied with input $x_1, \ldots, x_n$ and produce output $y$. In general, we assume that we deal with \emph{recursive} formal
systems, that is, the set of axioms is decidable and the set of all provable sentences (i.e., the set of all theorems) is \emph{recursively enumerable} or semi-decidable: if, given an arbitrary wff,  there is an algorithm which correctly determines when the formula is provable within the system, but may either produce  a negative answer or return no answer at all when the formula is not provable within the system.

Many important problems expressible in formal systems are undecidable, and this is captured in G\"odel's Incompleteness Theorems about any formal system with first-order logic (first-order predicate calculus) and containing Peano's axioms of arithmetic: (i) any such formal system is such that, if it is consistent, then it is incomplete: there are wff's which can neither be proved nor disproved; (ii) moreover, such a formal system cannot demonstrate its own consistency.

\subsubsection{Formal undecidability}

We shall discuss several essential steps required in a typical proof of G\"odel Incompleteness Theorems. Firstly, as we are dealing with arithmetic, we need to name, i.e., give a formal term (``numeral''), to each number: this is achieved by canonically denoting the natural number $n$ by numeral $\underline{n}$. Assuming that the primitive symbols, i.e. constant signs, such as `$0$' (zero) or '$S$' (denoting ``an immediate successor of ...'') \cite{Nagel2001,Buldt2016} are available (directly or via interpretation), the canonical
way to represent a natural number `$n$' in a formal system is via the numeral $\underline{n}$;
\[
\underline{n} \equiv \underbrace{S \ldots S}_{n \ \text{times}}  0 .
\]
One of the core insights of G\"odel was to encode the wff's of a formal system by natural numbers, by an ``arithmetization'', or ``G\"odel  numbering'', of the wff's.  Formally, for every wff $W$, the ``G\"odel  numbering'' scheme produces a natural number $\altmathcal{G}(W)$, i.e., the ``G\"odel  number'', which is further encoded by a numeral.  Such a code, the name of the ``G\"odel  number'' of a formula $W$, is denoted as $\ulcorner \ W \ \urcorner$.

To exemplify this, we firstly assign a natural number to each primitive symbol $s$ of the formal system (called the symbol number of $s$), e.g., symbol ``$0$'' is assigned number $1$ and symbol ``$=$'' is assigned number $5$.  Then we consider the wff $W$: ``0 = 0''. The G\"odel number for this formula is uniquely produced as the corresponding product of powers of consecutive prime numbers $(2, 3, 5, \ldots$), as $\altmathcal{G}(\text{``}0 = 0\text{''}) = 2^1 \times 3^5 \times 5^1 = 2 \times 243 \times 5 = 2430$. The name of the G\"odel number $\ulcorner \text{``}0 = 0\text{''} \urcorner$ is the numeral $\underline{2430}$.  Importantly, knowing $\altmathcal{G}(\text{``}0 = 0\text{''}) = 2430$ allows us to uniquely decode back into the wff's (due to the unique-prime-factorization theorem), by finding the unique sequence of prime factors, with associated exponents \cite{Raatikainen,Nagel2001}. 
G\"odel numbers are computable, and it is important to note that it is also effectively decidable whether a given number is a G\"odel number or not.  Formally, $\ulcorner \ W \ \urcorner$ is the numeral $\underline{\altmathcal{G}(W)}$, where $\altmathcal{G}(W)$ is the G\"odel number of $W$ \cite{Buldt2016,Gaifman2006naming}: 
\[
\ulcorner \ W \ \urcorner \equiv \underbrace{S \ldots\ldots S}_{\altmathcal{G}(W) \ \text{times}}  0 .
\]

One of the essential steps implicit in  G\"odel's proof is the Self-reference lemma \cite{Raatikainen,Smullyan1984}:
\begin{lemma}
Let $Q(x)$ be an arbitrary formula of formal system $\altmathcal{F}$ with only one free variable. Then there is a sentence (formula without free variables) $W$  such that
\[    \altmathcal{F} \vdash \ W \leftrightarrow Q(\ulcorner \ W \ \urcorner) \ . \]
\end{lemma}
This lemma is sometimes called the  Fixed-point lemma or the Diagonalization  lemma. This result was explicitly presented in 1934  by Carnap  \cite{Carnap1934}, phrased in different language, and was also used by Tarski in 1936 in proving the undefinability theorem: \emph{arithmetical truth cannot be defined in arithmetic} \cite{Tarski1936}.
The Self-reference Lemma establishes  that for any formula $Q(x)$ that describes a property of a numeral, there exists a sentence $W$ that is logically equivalent to the sentence $Q(\ulcorner \ W \ \urcorner)$.
 The arithmetical  formula $Q(x)$ describes a property of its argument, e.g., a numeral $x$, and hence, the expression $Q(\ulcorner \ W \ \urcorner)$ describes a property of the numeral  $\ulcorner \ W \ \urcorner$. This is the numeral of the G\"odel number of the formula $W$ itself.  Since the formula $W$  is logically equivalent to the formula $Q(\ulcorner \ W \ \urcorner)$, one can say that  the formula $W$ is referring to a property of  itself (being an argument of the right-hand side). 

Strictly speaking, as pointed out by \cite{Raatikainen}, the lemma only provides a (provable) material equivalence between $W$ and $Q(\ulcorner \ W \ \urcorner)$, and one should not claim ``any sort of sameness of meaning''.  It is, nevertheless, illustrative to consider a related result, a variant of the Mocking Bird Puzzle \cite{Smullyan1984}, which reflects the idea of the Lemma's proof and constructs a self-referential relation.  
\begin{quote}
``We are given a collection of birds. Given any birds $B$, $C$, if a spectator calls
out the name of $C$ to $B$, the bird $B$ responds by calling back the name of some
bird $B(C)$ (Thus each bird $B$ induces a function from birds to birds.) If
$B(C) = C$, then we say that $B$ is fixated on $C$. We call $B$ egocentric if $B$ is
fixated on itself. We are given that the set of functions induced by the birds is
closed under composition (more explicitly, for any birds $B$, $C$ there is a bird $D$
such that for every bird $X$, $D(X) = B(C(X))$ ). We are also given that there is a
bird $M$ (called a mocking bird) such that for every bird $B$, $M(B) = B(B)$. The
problem is to prove that every bird is fixated on at least one bird, and that at least
one bird is egocentric.''
\end{quote}
The proof has several instructive steps \cite{Smullyan1984}, reproduced here for convenience. Firstly, applying the closure under composition to a mocking bird $M$ we note that there must be a bird $D$ such that for every bird $X$, we have $D(X) = B(M(X))$. Then substituting $D$ for $X$, we obtain $D(D) = B(M(D))$. By definition of a mocking bird, $M(D) = D(D)$, and so we reduce to $D(D) = B(D(D))$, showing that bird $B$ is fixated on the bird $D(D)$, completing the first part (proving that every bird is fixated on at least one bird). Hence, the mocking bird must also be fixated on some bird $E$, that is, $M(E) = E$. Again, by definition of a mocking bird, $M(E) = E(E)$, yielding $E(E) = E$ and completing the second part (proving that at least one bird is egocentric).

Obviously, the substitutions in this proof were made simple by ignoring the encoding and decoding of birds as arguments but it is still interesting to note that the mocking bird can be seen as an analogy of universal computation (a universal Turing machine or a universal cellular automaton, capable of emulating computation of any other device, see Section \ref{univ}).

One now needs  to define the provability predicate $\textrm{Provable}_{\altmathcal{F}}(x)$ which captures the property of $x$ being provable in $\altmathcal{F}$.  Let the formula $\textrm{Proof}_{\altmathcal{F}}(y, x)$ strongly represent the binary relation ``$y$ is (the G\"odel number of) a proof of the formula (with the G\"odel number) $x$'' (following \cite{Raatikainen}, we note that it is always decidable whether a given sequence of formulas $y$ constitutes a proof of a given sentence $x$, according to the rules of the formal system ${\altmathcal{F}}$). The property of being provable in $\altmathcal{F}$ can then be defined as $\exists y \textrm{Proof}_{\altmathcal{F}}(y, x)$, abbreviated as $\textrm{Provable}_{\altmathcal{F}}(x)$.

The final step leading to G\"odel's First Incompleteness Theorem is an application of  the Self-reference lemma to the negated provability predicate $\neg \textrm{Provable}_{\altmathcal{F}}(x)$: 
\begin{equation}
\label{negProv}
    \altmathcal{F} \vdash \ W \leftrightarrow \neg \textrm{Provable}_{\altmathcal{F}}(\ulcorner \ W \ \urcorner) \ . 
\end{equation}
This then formally demonstrates that the system $\altmathcal{F}$ can derive that $W$ is true if and only if it is not provable in $\altmathcal{F}$. Furthermore, if the system $\altmathcal{F}$ is consistent, then it can be shown that the sentence $W$ is neither provable nor disprovable in $\altmathcal{F}$, showing the system to be incomplete.  It is important to point out that the G\"odel sentence $W$   can be constructed as a well-formed formula of the system $\altmathcal{F}$.

Common treatments of this seminal result interpret this theorem somewhat less formally, e.g., stating that the G\"odel sentence $W$  expresses or refers to its own unprovability \cite{Sieg2005}, analogous to  the Liar paradox: (the sentence claiming ``this sentence is false'' that can be neither true nor false). This can be traced back to the original G\"odel's work, where he  informally wrote: ``We therefore have before us a proposition that says about itself that it is not provable.'' \cite[p. 149]{godel2003collected}.   

What is important for our main purposes is that G\"odel's First Incompleteness Theorem can be used to demonstrate undecidability \cite{Raatikainen}. 
A formal system $\altmathcal{F}$ is  decidable if the set of its theorems is strongly representable in $\altmathcal{F}$ itself: there is some formula $\textrm{P}(x)$ of $\altmathcal{F}$ such that
\begin{equation}
\label{repr}
\begin{aligned}
 \altmathcal{F} \vdash \textrm{P}(\ulcorner \ W \ \urcorner) \ &\mathrm{whenever} \ \altmathcal{F} \vdash W, \ \mathrm{and}\\  
\altmathcal{F} \vdash \neg \textrm{P}(\ulcorner \ W \ \urcorner) \ &\mathrm{whenever} \ \altmathcal{F} \nvdash W \ . 
\end{aligned}
\end{equation}
For a weakly representable set of theorems only the first line of (\ref{repr}) is required (semi-decidability), that is, negations are not necessarily ``attributable'' to non-derivable formulas.
However,  it is possible to construct, within the system $\altmathcal{F}$, a G\"odel sentence $V^{\textrm{P}}$ relative to $\textrm{P}(x)$:
\begin{equation}
\label{phip}
   \altmathcal{F} \vdash \ V^{\textrm{P}} \leftrightarrow \neg \textrm{P}(\ulcorner \ V^{\textrm{P}} \ \urcorner) \ . 
\end{equation}
A contradiction follows, and hence, at least for this sentence the strong representability does not hold, and therefore, $\altmathcal{F}$ must be undecidable.
Crucially,  the G\"odel sentence $V^{\textrm{P}}$ is constructed as $V(\ulcorner \ V(x) \urcorner)$ for some wff $V(x)$ with a free variable, and so our central expression (\ref{phip}) explicitly states
\begin{equation}
\label{vv}
   \altmathcal{F} \vdash \ V(\ulcorner \ V(x) \urcorner) \leftrightarrow \neg \textrm{P}(\ulcorner \ V(\ulcorner \ V(x) \urcorner) \urcorner) \ . 
\end{equation}
This  perspective makes it explicit that the self-reference (or diagonalization) is used twice: inside and outside of the representative predicate $\textrm{P}(x)$, which is ``sandwiched'' between the two self-references \cite{Gaifman2007easy}.

The interrelationships played by fixed points, diagonalization, and self-reference in proofs of G\"odel's first incompleteness theorem are discussed in \cite{Buldt2016}, and we shall revisit these aspects in Section \ref{diag}.

\subsection{Turing Machines and The Halting Problem}
\label{TM}

Turing machines were introduced as a formal model of computation, intended as an abstract general-purpose computing device which modifies symbols on an infinite tape (\emph{data}) according to a finite set of rules (\emph{program}).  Prior to Turing's work the concept of an ``effective process'' had not been formalized, and so Turing's insight was to define the notion of an \emph{algorithm}: an automated process that is able to proceed, using a set of predefined rules, through a finite number of well-defined successive states, eventually terminating at a final state and producing an output. 

The infinite tape of a Turing machine (TM) is divided into discrete cells, thus implementing an unlimited memory capacity. The data are encoded, using some alphabet, as the initial input string, while the remaining cells on the tape contain blank symbols.
A TM employs a tape head which can move left and right across the tape, as well as read and write symbols contained in the cell to which the head points, thereby creating strings of symbols, from an \emph{alphabet} $\Gamma$, on the tape (a \emph{string} over an alphabet is defined as a finite sequence of symbols from that alphabet, while a \emph{language} over an alphabet  is defined as a set of strings \cite{Sipser}). 

These actions of the machine simulate an algorithm by following, at every given state, the rules described in its transition function, defined over a set of \emph{internal} states $Q$ and the alphabet $\Gamma$, as $\mu: Q \times \Gamma \rightarrow Q \times \Gamma \times \{L, R\}$. For example, if the machine is at a state $q_1$ and the tape head reads symbol $a$, then according to the machine's rules it may need to overwrite symbol $a$ with symbol $b$ on the tape, following which the machine switches its state to $q_2$  and moves to the right. Formally, this example can be expressed as $\mu(q_1, a) = (q_2, b, R)$.

The machine is able to distinguish certain predefined final states. For instance, if the machine enters the state $q_{acc} \in Q$, this indicates that the initial input is \emph{accepted} by the machine, while entering another state $q_{rej} \in Q$ represents that the input is \emph{rejected}. Both of these outcomes  cause the machine to halt, otherwise, the machine will continue its transitions forever \cite[pp. 138 -- 140]{Sipser}.

\subsubsection{Technical preliminaries}
\label{TM-tech}

A Turing machine, as adopted here following Sipser \cite[p. 140]{Sipser} and Hopcroft and Ullman \cite[p. 81]{Hopcroft1969}, is a tuple
\[ M = \langle Q, \Sigma, \Gamma, \mu, q_0, q_{acc}, q_{rej} \rangle \]
where $Q$, $\Sigma$ and $\Gamma$ are non-empty finite sets, and 
\begin{enumerate}
\item $Q$ is a set of states;
\item $q_0 \in Q$ is the start state;
\item $q_{acc} \in Q$ is the accept state;
\item $q_{rej} \in Q$ is the reject state;
\item $\Sigma$ is the input alphabet not containing the blank symbol $\vartextvisiblespace$;
\item $\Gamma$ is the tape alphabet, where  $\vartextvisiblespace \in \Gamma$ and $\Sigma \subseteq \Gamma \setminus \{\vartextvisiblespace\}$;
\item $\mu: Q \times \Gamma \rightarrow Q \times \Gamma \times \{L, R\}$ is a partial function called the transition function, where $L$ is left shift, and $R$ is right shift. If $\mu$ is not defined on the current state and the current tape symbol, then the machine halts.
\end{enumerate}
The transition function $\mu$ may be undefined for some arguments. Specifically, the machine $M$ halts in the accept $q_{acc}$ state (the initial tape contents is then said to be accepted by $M$) or the reject $q_{rej}$ state (the initial input tape is said to be rejected by $M$).  
With this definition, the \emph{output} of the computation, if it halts, is the determination whether the initial input is accepted or rejected.
However, one may equivalently define a TM with just one halting state $q_{halt} \in Q$, instead of two explicit accept and reject states.  In this case, if the machine halts, i.e. if it enters the state $q_{halt}$, then some content written on the same tape captures the actual output of the machine's computation.  The precise position of such output on the tape depends in general on some \emph{convention} and may be recognized in relation to the head position in a predesignated way, e.g., the head pointing to the cell containing the leftmost symbol of the output.  

This means that in the definition of a TM with two final  states $q_{acc}$ and $q_{rej}$, the initial tape input represents both some initial data and some target to be verified (to be either accepted or rejected): the final content written on the tape when the machine halts at either $q_{acc}$ or $q_{rej}$ does not matter.  On the contrary, in the alternative definition with just one halting state $q_{halt}$, the target is not included on the tape's initial input: instead it is expected to be found as the output on the tape when the machine halts.

In the first case, when the target is given within the input tape, the machine needs to only accept or reject this initial input. In the second case the final output needs to be explicitly generated  on the tape at the end of computation. 
Such flexibility in embedding the target reflects the duality of the data and the program in TMs, in the sense that a part of the input data may instead be represented in the internal  machinery, and vice versa. Technically, one may construct a TM working with an empty input tape, while solving a task completely embedded in the transition function over a certain set of internal states.
 
Given the current state $q$ and the current content on the tape in the form $uv$, where  two strings $u$ and $v$ are formed by symbols from $\Gamma$, with the head pointing to the first symbol of $v$, one may  define a \emph{configuration} of the TM  as $u \ q \ v$  \cite[p. 140]{Sipser}. For example, $11q_1011$ is the configuration when the tape is $11011$, the current state is $q_1$, and the head points to $0$. 

A Turing machine capable of simulating \emph{any} other TM is called a universal Turing machine (UTM) and provides a standard for comparison between various computational systems. In fact, the problems solvable by a UTM are exactly those problems solvable by an algorithm or any effective method of computation.

\subsubsection{Incomputability}
\label{TM:incomp}

A Turing machine $M$ recognises the language $L_M$ if and only if the set $L_M$ contains all the strings that machine $M$ accepts.
In demonstrating the Halting Problem for TMs, we will show, following Sipser \cite[p. 165]{Sipser}, the undecidability of the language
\[A_{TM}  = \{[M, w] \ | \ M \ \textrm{is a TM and} \ M \ \textrm{accepts the string} \ w \} \ , \]
where strings $w$ are formed by symbols from the alphabet $\Sigma$, that is, all strings in the set $\Sigma^*$ formed by the Kleene operator, and $[\cdot]$ denotes an \emph{encoding} of an object into a string using the alphabet $\Sigma$. Specifically, one may construct the encoding of a TM $M$, denoted $[M]$, into a regular string that comprises the \emph{description} of the tuple $M$. If needed, the string $[M]$ may  be  encoded in a binary regular form.   One may also encode compound objects, for example, create an encoding $[M, w]$ of two elements $M$ and $w$ together, as long as there is a way to interpret such an encoding as having two components. In terms of computability, $[\cdot]$ and its partial inverse (i.e., \emph{decoding}) must be effectively computable.

It will be crucial to deal with encodings $[M, [M]]$ so that such an input to another TM $P$ can be decoded into two components: the description of the machine $M$ and the input string $[M]$ into the machine $M$ itself.  The practical implementation of a decoding can vary, and one example (constructing, in fact, a universal TM simulating a machine $M$) separates the input data $[M]$ from the description of the machine $M$ by three consecutive $c$'s \cite[p. 102-104]{Hopcroft1969}, i.e., by a specific symbol sequence.

A typical approach to the proof of undecidability of language $A_{TM}$ involves an assumption that $A_{TM}$ is decidable leading to a contradiction. That is, we assume that there exists a decider TM $P$ (note the analogy with the representative predicate $\textrm{P}(x)$ used in the proof of undecidability of formal systems) such  that on input $[M, w]$, where $M$ is a TM and $w$ is a string, the decider $P$ halts and accepts $w$ if $M$ accepts $w$, while $P$ halts and rejects $w$ if $M$ fails to accept $w$.   Formally, the decider machine $P$ is defined as
\begin{equation}
\label{t:p}
    P([M, w]) = 
\begin{cases}
accept \hspace{7mm} \textrm{if} \ M \ \textrm{accepts} \ w    \\%[0.8em]
reject \hspace{7mm}  \textrm{if} \ M \ \textrm{does not accept} \ w                     
\end{cases}
\end{equation}
As an aside, the decider machine $P$ is not a UTM that can simulate an arbitrary TM on arbitrary input. Unlike the decider $P$ which rejects when $M$ loops on $w$, a UTM simulating $M$ would run forever on $w$ if $M$ runs forever on $w$. It is the assumed decidability of the universal decider $P$ which will be refuted in the proof.

Then we construct another machine $V$ that is able to (i) interpret its input $[M]$ as the encoding of some TM $M$,  (ii) invoke, as a subroutine, the decider machine $P$ with input $[M, [M]]$, and (iii) once the decider  $P$ halts with either accept or reject (which is ensured by the assumption that $P$ must halt on any input $[M, w]$), the machine $V$ inverts the outcome of $P$. That is, the machine $V$ accepts the input $[M]$ if $P([M, [M]])$ rejects its compound input (which happens, by definition of $P$, if $M$ does not accept $[M]$), and  rejects if $P([M, [M]])$ accepts (that is, if $M$ accepts $[M]$). Formally, the inverter machine $V$, which includes three distinct steps, is defined as follows:
\begin{equation}
\label{t:d}
    V([M]) = 
\begin{cases}
reject \hspace{7mm} \textrm{if} \ M \ \textrm{accepts} \ [M]    \\%[0.8em]
accept \hspace{7mm}  \textrm{if} \ M \ \textrm{does not accept} \ [M]                     
\end{cases}
\end{equation}
In creating the input $[M, [M]]$ for the decider machine $P$ we forced the machine $M$ to run on the input representing its own description $[M]$. This is a manifestation of self-reference (similar to the ``inside'' self-reference used in construction of the G\"odel sentence). 

It is also important to realise that the input to the inverter machine $V$ is given by the encoding $[M]$ and not by the compound object $[M, [M]]$ which is constructed by $V$ before calling the ``sandwiched'' decider subroutine $P$. This construction is possible because both the encoding and decoding are effectively computable (again we draw an analogy with the encoding and decoding utilized by G\"odel numbering scheme $\ulcorner \ W \ \urcorner = \underline{\altmathcal{G}(W)}$).   

The final step is to run the inverter machine $V$ on itself, that is, to consider $V([V])$ (in analogy to the ``outside'' self-reference in G\"odel's proof):
\begin{equation}
\label{t:dd}
    V([V]) = 
\begin{cases}
reject \hspace{7mm} \textrm{if} \ V \ \textrm{accepts} \ [V]    \\%[0.8em]
accept \hspace{7mm}  \textrm{if} \ V \ \textrm{does not accept} \ [V]                     
\end{cases}
\end{equation}
This is, of course, a contradiction analogous to the Liar's Paradox (or the inconsistency shown by the G\"odel sentence in formal systems): the inverter machine $V$ rejects its input $[V]$ whenever $V$ accepts
$[V]$. This contradiction shows the impossibility of the decider TM $P$, and hence, the undecidability of language $A_{TM}$. 
One corollary is that the language $A_{TM}$ is TM recognisable but not  decidable

As the proof shows, the undecidability arises due to the self-referential ability of a TM to interpret and run an input which encodes its own description, reflecting the program-data duality.
The program-data duality, allowing programs to interpret other programs (sets of rules) as data (encoded strings), makes it possible for TMs to answer questions about, and ultimately completely emulate, the behaviour of other TMs. 
It is this implicit self-referential ability that results from the program-data duality that leads to the undecidability and The Halting Problem.

\subsection{Cellular Automata and The Edge of Chaos}
\label{CA}
\vspace*{2mm}

\subsubsection{Technical preliminaries}

A Cellular Automaton (CA) is a discrete dynamical system $C$ \cite{Wolfram1984} defined on a $d$-dimensional lattice $c$. Each lattice site (cell) $c_i$ takes a value from a finite alphabet $A_C$, i.e., $c_i \in A_C$, where the indexing reflects the dimensionality and geometry of the lattice \cite{goldenfeld}. For example, for a $1$-dimensional CA ($d = 1$), index $i \in \mathbb{Z}$, the set of integers.  A configuration $c$ of cells in the lattice is a bi-infinite sequence of specific cell values $c_i$, that is, $c = (\ldots, c_{-2}, c_{-1}, c_{0}, c_1, c_2, \ldots)$, for instance, in a $1$-dimensional CA with a binary alphabet $A_C = \{0,1\}$ a configuration may look like $(\ldots, 0, 1, 1, 0, 1, \ldots)$. Most applied work with CAs considers finite automata, but infinity is necessary to generate undecidable dynamics, similarly to the infinite tape in TMs.

Each cell is updated in discrete time steps $t$ according to a deterministic \emph{local} rule $\phi_C$ involving values of $r$ neighbouring cells, and by convention a cell is included in its neighbourhood:
\begin{equation}
\phi_C: A_C^{(2r+1)^d} \rightarrow A_C
\end{equation}
so that the value of the $i$-th cell at time $t$ is updated as follows:
\begin{equation}
c_i^t = \phi_C(c_{i-r}^{t-1}, c_{i-r+1}^{t-1}, \ldots, c_{i+r}^{t-1}) 
\end{equation}
The set of all configurations will be denoted as $\Psi_C = A_C^{\mathbb{Z}^d}$.
This local rule yields a global mapping (global rule) setting temporal dynamics on the lattice:
\begin{equation}
\Phi_C: \Psi_C \rightarrow \Psi_C
\end{equation}
The configuration $c$ at time $t$ is completely determined by the preceding configuration: 
\begin{equation}
c^t = \Phi_C(c^{t-1}) \ ,
\end{equation}
while the initial configuration $c^0$ is a sequence of cells in the lattice at time $t = 0$.

Formally, a CA $C$  is a tuple:
\begin{equation}
C = \langle  A_C, d, \phi_C \rangle \ ,
\end{equation}
and in order to specify its dynamics  we shall use the notation $C(c^0)$ for the initial configuration $c^0$.

For example, a one-dimensional ($d =  1$) CA $C$ with a binary alphabet $A_C =\{0, 1\}$ may use a local update rule $\phi_C$ defined for a neighbourhood with 3 cells (i.e., $r = 1$), setting dynamic updates as:
\begin{equation}
c_i^t = \phi_C(c_{i-1}^{t-1}, c_{i}^{t-1}, c_{i+1}^{t-1}) 
\end{equation}
There are $8 = 2^{3}$ permutations of inputs  into the local rule $\phi_C$, and consequently, $256 = 2^8$ local rules in total. This type of CA with two possible values for each cell and local update rules defined only on the current state of the cell and its two nearest neighbors is called Elementary Cellular Automata (ECAs).  A scheme, known as the Wolfram code, assigns each ECA rule a number from 0 to 255 as follows: the resulting states for each possible input permutation (written in order $111, 110, \ldots, 001, 000$) is interpreted as the binary representation of an integer. For instance, the ordered resulting states $0,1,1,0,1,1,1,0$ of $\phi_C$ constitute the rule $110$, because the integer $110$ has a binary representation of $01101110$ \cite{Wolfram2002}. 
The rule $110$ is of particular interest because it is the only one-dimensional  CA which has been proven to have the same computational power as a UTM \cite{Cook2004}, and therefore, can generate undecidable dynamics.

Another well-studied example is \emph{Conway's Game of Life} \cite{Gardner1970}: a two-dimensional ($d =  2$) CA $G$ with a binary alphabet $A_G =\{0, 1\}$ and a specific local update rule $\phi_G$ defined for  the Moore neighbourhood with 9 cells (i.e., $r = 1$): $\phi_G: A_G^{3^2} \rightarrow A_G$, such that
\begin{enumerate}
\item  Deaths. Any live cell with fewer than two or more  than three live neighbours dies.
\item  Survivals. Any live cell with two or three live neighbours lives on to the next generation.
\item  Births. Any dead cell with exactly three live neighbours becomes a live cell.
\end{enumerate}
It is well-known that Game of Life is also undecidable,  having the same computational power as a universal Turing machine  \cite{Berlekamp1982}. 

\begin{figure}
\begin{center}
\includegraphics[width=1.0\columnwidth]{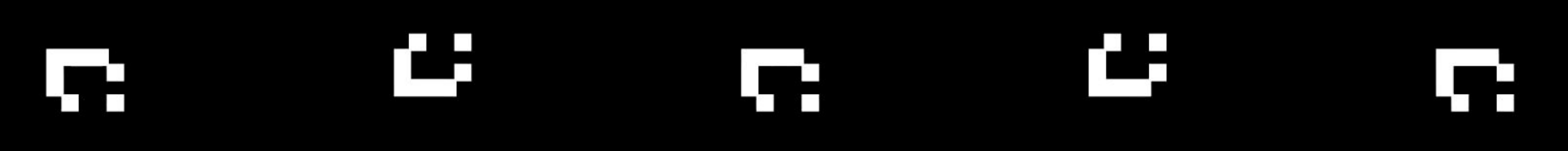}
\caption{The Game of Life: gliders (``lightweight spaceships'' (LWSS)). Snapshot of dynamics ``Life in Life'' by Phillip Bradbury: \url{https://www.youtube.com/watch?time_continue=4&v=xP5-iIeKXE8}, used under CC BY license.}
\label{fig:gol1}
\end{center}
\end{figure}

Both one-dimensional rule $110$ and two-dimensional Game of Life produce gliders: coherent spatial patterns that move across the grid replicating their structure (see Fig. \ref{fig:gol1}). It has been demonstrated that gliders fulfill the role of information transfer in distributed computation carried out by CA \cite{liz10d}.

\subsubsection{Termination condition}

As our general purpose is to study analogies and equivalencies between CAs and other systems which compute or prove specific outcomes, we need to adopt a convention determining when the desired output has occurred, i.e., trace the dynamics of the input configuration until some
``halting'' condition applies \cite{Sutner-chapter}. For example, the end of computation may be indicated by  reaching  an (attractive) fixed-point or by reaching a temporal cycle of length two: this can be determined by comparing configurations at different time steps \cite{Lindgren1990}. Importantly, as pointed out by Sutner \cite{Sutner-chapter}, this condition must be primitive recursively decidable, but a precise mechanism may vary: for example, a termination condition may check if a particular predesignated cell reaches a special state, or if an arbitrary cell or a set of cells reach a special predefined state(s), or if the configuration is a fixed point  or a limit cycle. 

Importantly, we distinguish among  attractors, i.e. limit cycles (including fixed points which are limit cycles of length 1) $c^*$ by arbitrarily designating some of those as ``accepted'' and the rest as ``rejected'' outcomes (to stay closer to the intuition behind the proof of undecidability for TMs presented in section \ref{TM:incomp}). Illustrating this for fixed points, this  can be done by arbitrarily partitioning the set of all configurations $\Psi_C$ into two sets, $\Psi_C^+$ and $\Psi_C^- = \Psi_C \setminus \Psi_C^+$, so that an attractive fixed-point configuration $c^t \in \Psi_C^+$ can be interpreted as an accepted outcome, and a fixed-point configuration $c^t \in \Psi_C^-$ would correspond to a rejected outcome.  This partitioning is formally described by function $\pi_C: \Psi_C \times \Psi_C \times \mathbb{N} \rightarrow \{1,0\}$ such that, at time $t \in \mathbb{N}$, $\pi_C(c^t) = 1$ if and only if $c^t = c^{t-1}$, and $c^t \in \Psi_C^+$, while $\pi_C(c^t) = 0$ if and only if $c^t = c^{t-1}$ and $c^t \in \Psi_C^-$. 
In order to make concrete this arbitrary partition of the configuration space, we may choose any single cell, e.g., $c^t_{42}$, then select a specific symbol $\alpha \in A_C$, and then, for a fixed-point $c^t$, assign  $\pi_C(c^t) = 1$ if and only if $c^t_{42} = \alpha$, and  $\pi_C(c^t) = 0$ if and only if $c^t_{42} \neq \alpha$. 
 
 In demonstrating that rule $110$ is computationally equivalent to a UTM, Cook developed a concrete algorithm for compiling a Turing machine showing that the dynamics of rule $110$ will eventually produce the bit sequence $01101001101000$ if and only if the corresponding Turing machine halts \cite{Cook2008}.  
Such a termination condition can be expressed in terms of temporal rather than spatial sequences: ``it is also the case that the sequence $110101010111111$ will be produced over time by a single cell if and only if the Turing machine halts'' \cite{Cook2008}. Specifically, these sequences are produced by a designated glider configuration (glider $F$), chosen to occur only if the corresponding algorithm halts.  Similarly, one may designate appearance of a specific two-dimensional configuration in the Game of Life --- glider, still-life (a non-changing pattern), or oscillator (a pattern returning to its original state, in the same orientation and position, after a finite number of generations) --- as the ``accepted'' termination condition. Analogously, another glider, still-life, or oscillator configuration may be chosen to indicate the opposite ``rejected'' termination outcome.  We stress that, in order to achieve the computational equivalence with Turing machines, such termination conditions are necessary to specify in addition to setting the automaton's rule table and an initial configuration.

Therefore, in general, one may extend the definition of a CA $C$ to include a termination condition $\pi_C$: 
\begin{equation}
C = \langle  A_C, d, \phi_C, \pi_C \rangle
\end{equation}
 so that $C(c^0)$ specifies the CA dynamics starting from initial configuration $c^0$.

The inequality $c^t \neq c^{t-1}$ is always computable.  However, due to the finitary nature of all computations, the equality is not decidable in type-2 computability \cite{Sutner-chapter} (the framework of Type-2 Theory of Effectivity  allows for computability over sets of a cardinality up to continuum \cite{weihrauch2000computable}), and so there is no guarantee that the termination condition can be effectively checked for any given pair $c^t$ and $c^{t-1}$, because the lattice is itself infinite. 
As we shall see in subsection \ref{CA:class}, one may restrict the space of possible CA configurations to certain subspaces within which the termination condition can always be checked in a primitive recursively decidable manner. 
Henceforth we follow the approach which restricts the space of possible CA configurations to only those subspaces over which a recursively decidable test of termination conditions is possible. As pointed out by Sutner \cite{Sutner-chapter}, all of these subspaces are closed under the application of a global map $\Phi_C$, ensuring that the dynamics stay within the restricted space.

We will abbreviate the case when a CA $C$ terminates at a configuration $c^t \in \Psi_C^+$ as follows $C: c^0 \rightarrow c^+$, and the case terminating at $c^t \in \Psi_C^-$ as  $C: c^0 \rightarrow c^-$.  It is worth pointing out that membership  $c^t \in \Psi_C^+$ or  $c^t \in \Psi_C^-$ is computable within the restricted subspace of possible CA configurations, i.e., a recursively decidable test of membership is ensured.

Our choice of the distinction between the  attractors in $\Psi_C^+$ and $\Psi_C^-$ as opposite outcomes of the computation carried by the dynamics is somewhat arbitrary. Importantly, any such distinction needs to be encodable into a regular string, for example, the determination  that an attractor satisfies the requirement of being effectively computable during CA run-time (i.e., it is intrinsic to CA dynamics), and all that needs to be encoded is the assignment of ``accept'' or ``reject'' labels to the chosen binary outcomes.

With such a termination condition it is possible to frame a question on decidability of CA dynamics directly, without tasking an algorithm external to the CA to check whether the CA dynamics do or do not ever reach the given target configuration.  

It is known that a TM $M$ can be simulated with a one-dimensional CA $C$, by creating the alphabet $A_C$ as the union of the set of states $Q$ and the tape alphabet $\Gamma$ of $M$, and constructing the local update rule $\phi_C$ out of the transition function $\mu$  by smartly interleaving state symbols $q \in Q$ and tape symbols $\gamma \in \Gamma$ \cite[p.~121]{Durand}.  For example, the transition resulting in the move of the machine's head to the right corresponds to these two local CA updates by $\phi_C$:
\[\text{if  } \mu(q_1, \gamma_1)=(q_2, \gamma_2, R), \ \ \text{then} \ \ \phi_C(*, *, q_1, \gamma_1, *) = \gamma_2 \ \ \text{and} \ \ \phi_C(*, q_1, \gamma_1, *, *) = q_2 \ ,\]
where * matches any state. One may see a parallel here with one-dimensional configurations of  TMs \ref{TM-tech}.
As a result, the computation carried out by a TM, updating over the set of states $Q$ and the tape alphabet $\Gamma$, i.e. over $Q \times \Gamma$, can be made equivalent to dynamics of the corresponding automaton which modifies its configurations $c^t \in A_C^{\mathbb{Z}}$. 
Consequently, the combination of the TM's start state $q_0 \in Q$ and its initial tape pattern formed by symbols from $\Sigma$ corresponds to the initial configuration $c^0 \in A_C^{\mathbb{Z}}$ of the CA. 

Finally, the role of the machine's accept and reject states $q_{acc} \in Q$ and $q_{rej} \in Q$ may be played by the termination condition $\pi_C$  checking whether configurations are  attractors in $\Psi_C^+$ or $\Psi_C^-$.

\subsubsection{Classifications of Cellular Automata}
\label{CA:class}

The repeated application of a global rule $\Phi_C$, starting from the initial configuration $c^0$, produces an evolution of configurations $c^t$ over time. 
In classifying global CA rules according to its long-term  asymptotic dynamics, the following qualitative taxonomy is typically employed \cite{Wolfram1984-b}: 
\begin{itemize}
\item class I (evolution leads to a homogeneous state); 
\item class II (evolution leads to periodic configurations); 
\item class III (evolution leads to chaotic patterns); 
\item class IV (evolution leads to complex localized structures over long transients).
\end{itemize}
In other words, class I consists of CAs that, after a finite number of time steps, produce a unique, homogeneous state (analogous to ``fixed point'' dynamics). Class II contains automata which generate a set of either stable or periodic structures (typically having small periods --- analogous to ``limit cycle'' dynamics) --- each region of the final configuration depends only on a finite region of the initial configuration.
Class III includes CAs producing aperiodic (``chaotic'') spatiotemporal patterns from almost all possible initial states --- the effects of changes in the initial configuration almost always propagate forever,  and a particular region of the final configuration depends on a region of the initial configuration of an ever-increasing size (analogous to ``chaotic attractors''). Class IV includes CAs that generate patterns continuously changing over an unbounded transient, and some of these CAs have been shown to be capable of universal computation \cite{Wolfram1984-b,Cook2004,Wolfram2002}. 

It is important to distinguish between (i) (possibly undecidable) questions about CA dynamics on all possible initial configurations, and therefore, about the CAs classification, and (ii) (possibly undecidable) questions whether the CA dynamics can ever reach a target configuration for a given initial configuration.  An extensive analysis of the classification problem and its undecidability for a broad range of cases has been provided by Sutner \cite{Sutner,Sutner-chapter} and others \cite{Durand, Kari}.  The important insight in dealing with the classification problem is a restriction of the space of possible configurations to certain subspaces, which include, for example, configurations with finite support, or spatially periodic configurations, or almost periodic configuration, or in the most general case,  recursive configurations, where a cell state is assigned by a computable function, so that such a restriction produces an ``effective dynamical system'' \cite{Sutner-chapter}.

\subsubsection{Universal Cellular Automata}
\label{univ}

A universal CA is a CA which can emulate any CA. One of the simplest universal CAs has been shown to be the rule 110 ECA with just 2 states which happen to be sufficient for producing universality in a 1-dimensional CA \cite{Cook2004}. 
A universal CA has the same power as a UTM, and can, therefore, generate undecidable dynamics. 
For example, whether an initial state will ever reach a quiescent state can be seen as  the CAs equivalent of the undecidable
Halting Problem \cite{Brady-Herken,goldenfeld}. 
The undecidability of CA dynamics and the role played by self-reference will be discussed in subsection \ref{CA:undecid}, and here we  point out several aspects that are important in constructing universal CAs.

First of all, in constructing universal CAs one must derive a way to encode any simulated CA and its initial configuration, as data, in the form that can be used by the universal CA. Without loss of generality, we can assume that such an encoding $[C,c^0]$ can be produced in  a primitive recursively decidable way, as one only needs to encode the initial configuration $c^0$ from the suitably restricted subspace (e.g., recursive configurations) and the local rule $\phi_C$ defined for finite neighbourhoods. %As mentioned earlier, the ability to determine whether a configuration is a fixed-point or not is intrinsic to CA and, hence, does not need to be encoded. Therefore, 
The encoding of a CA which has been extended with a termination condition $\pi_C$ needs only to include in addition the distinction between  attractors in $\Psi_C^+$ and $\Psi_C^-$. Such a distinction can be determined by the state of a single designated cell.

%Secondly, there are circumstances when the computational process being simulated contains several distinct sub-processes, which may interact with each other.  For example, the computational process carried out by the inverter TM $D$, described by (\ref{t:d}), is comprised of three distinct sequential steps. A single CA $C$ emulating such a machine could be constructed as a \emph{composition} of a number of sub-process CAs $C_i$, $1 \le i \le n$, denoted $C = \oplus C_i$, which can interpret and perform operations on the results of the intermediate dynamics \cite{Bandman2010}. The compositional techniques may vary, but typically involve an extended alphabet with additional symbols which are  used in selecting, at different stages, the update rules $\phi_{C_i}$ of specific sub-process CAs $C_i$ \cite[Def. 4.1]{Baldwin}, or combining the update rules into a compositional update $\phi_C$ \cite[Def. 17]{Fujio}. 

 Another technique employed in simulating CAs uses the coarse-graining of the CA dynamics, by grouping neighboring cells into a \emph{supercell} according to some specified convention  (this essentially follows a renormalization scheme) \cite{goldenfeld}. A supercell is created by projecting the states of a block of cells of one CA $C$ into a single cell of the coarse-grained CA $C'$.  
The update rule $\phi_{C'}$ is constructed from the update of $\phi_C$ by projecting its arguments and outcome, subject to certain commutativity conditions \cite{goldenfeld}, to the arguments and outcomes defined for supercells. Such a coarse-grained emulation of $C$  may or may not be carried out without loss of relevant dynamic information, but  a universal coarse-grained CA $C'$ ensures that all dynamics can be preserved.

\begin{figure}
\begin{center}
\includegraphics[width=1.0\columnwidth]{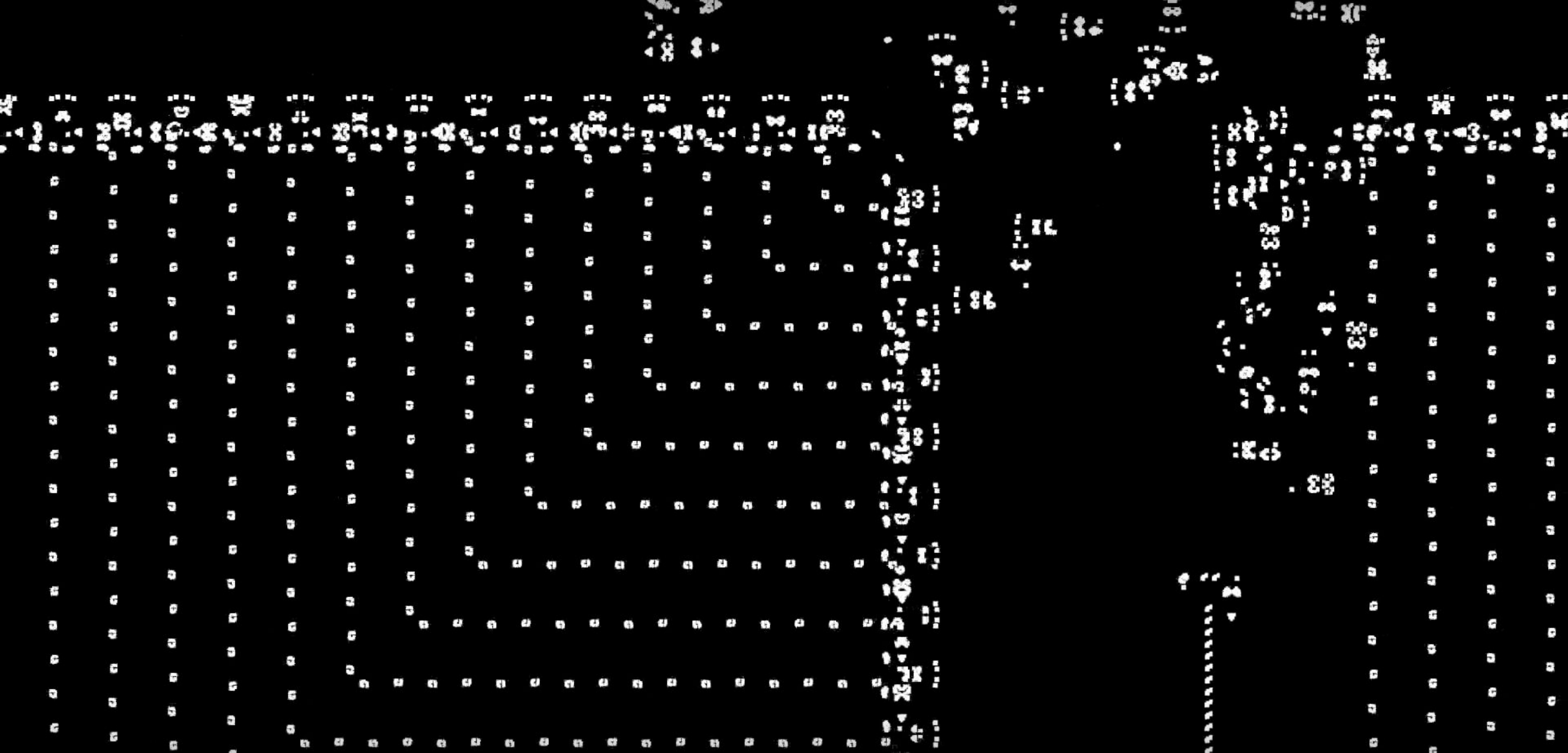}
\caption{The Game of Life simulated in OCTA metapixel: emergence of meta-level states within unit cells, which are being filled by a series of gliders formed by the underlying dynamics. Snapshot of dynamics ``Life in Life'' by Phillip Bradbury: \url{https://www.youtube.com/watch?time_continue=4&v=xP5-iIeKXE8}, used under CC BY license.}
\label{fig:gol2}
\end{center}
\end{figure}

\begin{figure}
\begin{center}
\includegraphics[width=1.0\columnwidth]{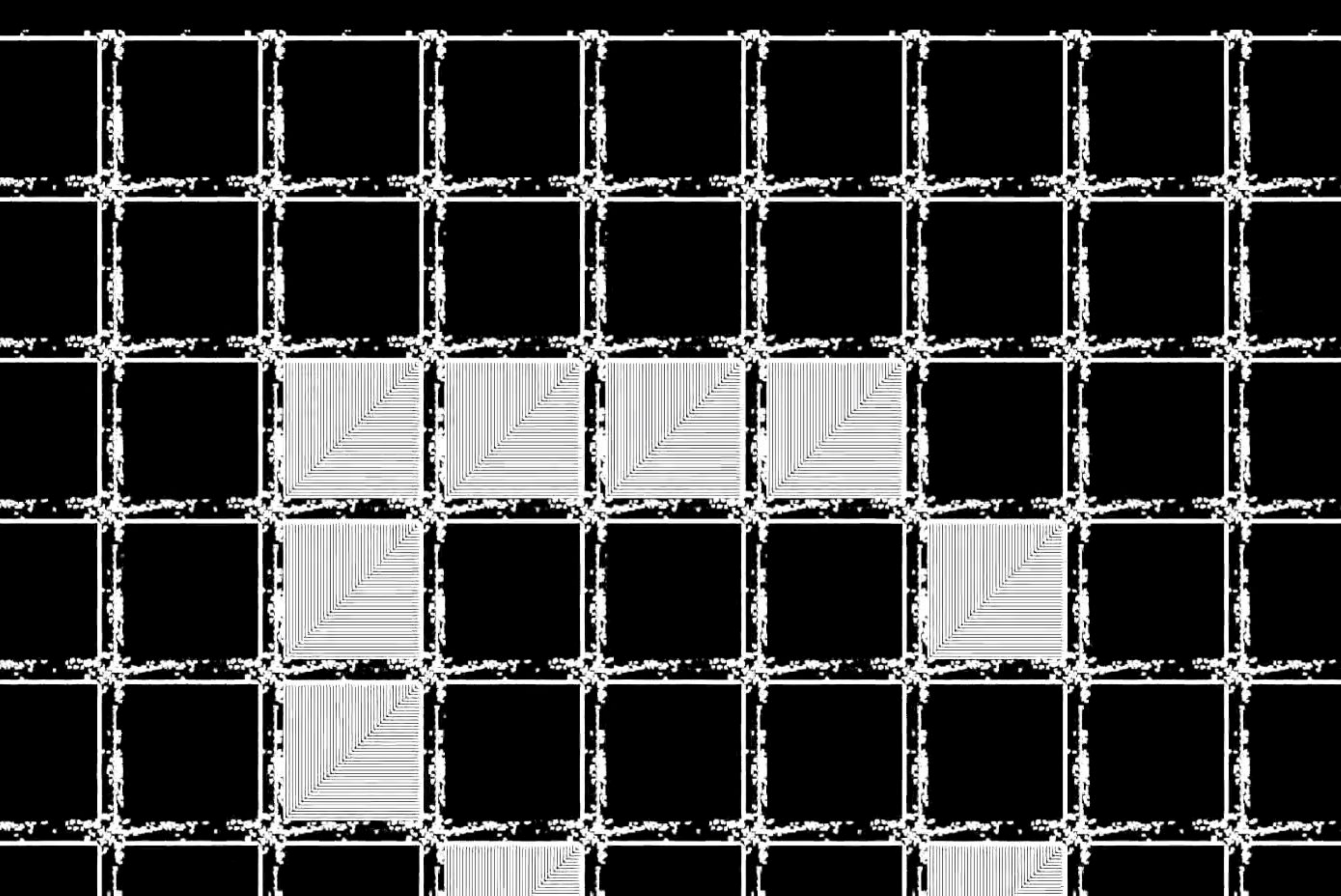}
\caption{The Game of Life simulated in OCTA metapixel: emergence of a meta-level LWSS glider configuration. Snapshot of dynamics ``Life in Life'' by Phillip Bradbury: \url{https://www.youtube.com/watch?time_continue=4&v=xP5-iIeKXE8}, used under CC BY license.}
\label{fig:gol3}
\end{center}
\end{figure}

An important building block used in constructing universal two-dimensional CAs is a \emph{unit cell}: a rectangular or square subset of the configuration space (e.g., the Game of Life plane) that tiles over the space. In general, a unit cell has a fixed number of distinct patterns, essentially forming a meta-level alphabet --- for example, two distinct patterns, the ON and OFF cells, are needed to simulate the binary Game of Life.  Each tile can assume one of the patterns, aiming to simulate a cellular automaton in a coarse-grained but fully preserving way. For example, the \emph{Outer Totalistic Cellular Automata metapixel} (\emph{OCTA metapixel}), a $2048 \times 2048$ unit cell, was designed by Brice Due in 2006 to reproduce the Game of Life and any Life-like CA \cite{wiki:octa} in a ``Life in Life'' simulation.  The period of OCTA metapixel is	35328 cycles, needed to change between the ON and OFF metapixel states (see Fig. \ref{fig:gol2} showing emergence of meta-level states during one period). The meta-level ON and OFF cells are particularly easy to distinguish  in a simulation of the Game of Life  by OCTA metapixel, as shown in Fig. \ref{fig:gol3}. 

Importantly, the unit cell's states, observed at the meta-level,  emerge as a result of the dynamics produced by the underlying CA, and not by any direct interaction between metapixels. That is, the distributed computation itself is still carried out at the underlying level (e.g., the level of the original Game of Life), but the ``Life in Life'' dynamics, which are recognized with respect to the OCTA metapixels' states, are simulated at the emergent level. The emergence here is understood not only as pattern formation, but also in the broader sense related to the efficiency of prediction \cite{pro09}. 

\begin{figure}
\begin{center}
\includegraphics[width=1.0\columnwidth]{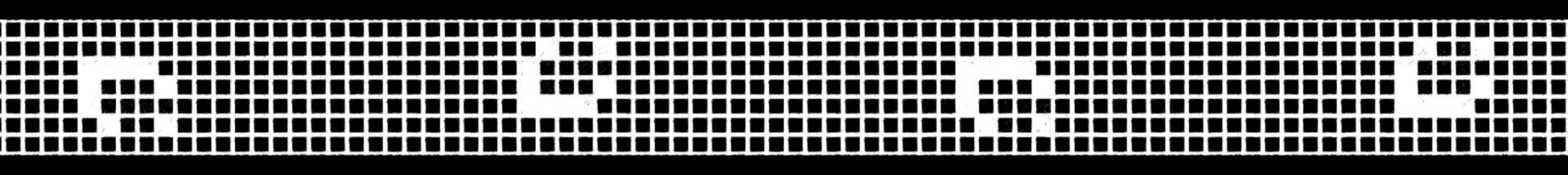}
\caption{The Game of Life simulated in OCTA metapixel: emergence of meta-level LWSS gliders. Snapshot of dynamics ``Life in Life'' by Phillip Bradbury: \url{https://www.youtube.com/watch?time_continue=4&v=xP5-iIeKXE8}, used under CC BY license.}
\label{fig:gol4}
\end{center}
\end{figure}

The dynamics of the underlying universal CA simulate the  ``Life in Life'' CA, completely reconstructing itself at the meta-level: see, for example, the emerging glider configuration shown by Fig. \ref{fig:gol3}, and a series of gliders shown in Fig. \ref{fig:gol4}.  Therefore, any  termination condition specified at the underlying level may also be utilized at the meta-level, with respect to the emergent pattern(s) defined in terms of metapixel states.

\subsubsection{Undecidable dynamics}
\label{CA:undecid}

In this subsection we sketch a proof of the undecidability of CA dynamics, following the steps used in the proof of the undecidability of language $A_{TM}$, which demonstrated the Halting Problem for TMs, as well as the undecidability of formal systems.  
The traditional approaches typically establish an equivalence between CAs and TMs \emph{per se}, and thus impute universality and undecidability of CAs based on these properties in TMs. 
Our purpose is more specific insofar as we aim to explicitly reconstruct the Halting Problem within the undecidable dynamics of CAs, exposing the Liar's Paradox analogy within this framework.

If CA dynamics were decidable, then there would have existed a  decider CA with a binary alphabet $P = \langle  (A_P = \{0, 1\}), d,  \phi_P, \pi_P \rangle$   capable of simulating any other CA $M = \langle  A_M, d,  \phi_M, \pi_M \rangle$ starting from the initial configuration $m^0$ (again we note the analogy with the representative predicate $\textrm{P}(x)$ used in the proof of undecidability of formal systems, and the decider TM $P$). As we have seen, a universal CA with a binary alphabet can be constructed, and it is the decidability of the dynamics created by a universal CA which we shall refute in the proof. The input of decider $P$ is given as $p^0 = [M,m^0]$, while the termination condition  $\pi_P$ are specified in such a way that only two decidable outcomes are possible, being constrained as follows:
\begin{equation}
\label{ca-h}
\begin{cases}
P: p^0 \rightarrow p^+  \hspace{7mm} \textrm{whenever} \ M: m^0 \rightarrow m^+    \\%[0.8em]
P: p^0 \rightarrow p^-  \hspace{7mm} \textrm{whenever} \ M: m^0 \rightarrow m^- \  \textrm{or runs forever} 
\end{cases}
\end{equation}
In other words, the  dynamics of $P$ terminate at some  attractor configuration $p^t$ whenever the dynamics of $M$ terminate at some  attractor configuration $m^t$.   More importantly, whenever the  dynamics of $M$ reach  an attractor in the complement set $\Psi_M^-$ or simply run forever, the  dynamics of $P$ are assumed to necessarily reach  an attractor in the complement set $\Psi_P^-$. The ability to specify such a definitive termination condition for $P$ is, in fact, the main assumption behind the decidability of CA dynamics, to be refuted by the proof that follows.   

 The universal CA $P$ that we shall use to illustrate the proof is the ``Life in Life'' CA, based on the OCTA metapixel.  As mentioned already, this CA is universal and the aspect to be refuted is the decidability of the dynamics created by the ``Life in Life'' CA --- in other words, we shall show that this CA is not a decider CA. In doing so, we specify the termination condition $\pi_P$ for the ``Life in Life'' CA, set to capture the two decidable outcomes \ref{ca-h}, in a way replicating the termination condition  $\pi_M$ of the  CA $M$, but expressed in the alphabet of the CA $P$. For example, the termination condition may be set with respect to observing specific Game of Life configurations, i.e., if  a designated oscillator configuration, $F^{+}$, is observed at the meta-level within the lattice configuration $c^t$, then $\pi_P(c^t) = 1$, while appearance of another specifically chosen oscillator configuration $F^{-}$, or the determination that the  CA $M$ runs forever, would yield $\pi_P(c^t) = 0$. Since, by the to-be-refuted assumption, $P$ is a decider CA, both of these outcomes must be decidable.

Having assumed that $P$ exists, we construct another inverter CA $V = \langle  A_V, d, \phi_V,  \pi_V \rangle$, running from the initial configuration $v^0 = [M]$.  This intends to match the idea of a G\"odel sentence in formal systems, as well as the inverter TM $V$. Using suitable encoding and decoding in   producing $[M, [M]]$ from $[M]$ is the first required step. For example, in simulating ``Life in Life'', the initial configuration $v^0$ of CA $V$ must match the initial configuration $m^0$ of the  CA $M$, and hence, must be encoded in a way ensuring that the initial metapixels form the ON and OFF states identical to the binary states of the initial configuration $m^0$.  
%That is, the  initial configuration $[M]$ does not immediately encode the compound object $[M, [M]]$ which could have been used as an initial configuration by the decider CA $P$.  However, it is possible to specify the updates of the local rule $\phi_V$  in such a way that the dynamics of $V$ emulate the dynamics of ``sandwiched'' $P$ running on the configuration $[M, [M]]$. This corresponds to the step (\ref{t:d}) in the undecidability proof for  TMs where the decoding from $[M]$ was used to create an encoding of $[M, [M]]$.  
Similar to the inverter TM $V$ described by (\ref{t:d}), the CA $V$ will 
%be a compositional CA $V =\oplus V_i$, accounting for (i) decoding of $[M]$ by some CA $V_1$, and (ii) 
simulate $M$ running on $[M]$. % by  CA $V_2 = P$.
  %In general,  ``copying'' from $[M]$  to $[M, [M]]$ required by $V_1$ can be achieved by gliders. 
The crucial step in creating the inverter CA is, however,  the inversion of  the  attractor outcomes, so that the termination condition $\pi_V$ matches the following:
\begin{equation}
\label{eqCA:V}
\begin{cases}
V: [M] \rightarrow v^-  \hspace{7mm} \textrm{whenever} \ M: [M] \rightarrow m^+    \\%[0.8em]
V: [M] \rightarrow v^+  \hspace{7mm} \textrm{whenever} \ M: [M] \rightarrow m^- \  \textrm{or runs forever} 
\end{cases}
\end{equation}
 It is important to point out that this inversion occurs by simply changing the interpretation of the  Game of Life configurations designated to indicate the termination outcomes. Formally, if the designated oscillator configuration $F^{-}$ is observed at the meta-level within the lattice configuration $c^t$, or it is determined that the  CA $M$ runs forever,  then $\pi_V(c^t) = 1$. On the contrary, if the designated oscillator configuration $F^{+}$ is observed at the meta-level within the configuration $c^t$, then $\pi_V(c^t) = 0$.  We stress  that the inversion of the termination conditions is confined to (re-)setting  $\pi_V$, outside of the specifications of the CA's rule table and initial configuration. Thus, the ``Life in Life'' CAs $P$ and $V$ simulate the  CA $M$ in exactly the same way, but the interpretations of the observed oscillators $F^{+}$ and $F^{-}$ are  inverted in $V$.
%The composition $V =\oplus V_i$ may be achieved in different ways, for example, by constructing and inverting the configuration $[M, [M]]$ within the decoding dynamics of $V_1$, while also inverting the actual dynamics of $V_2 = P$ (i.e., inverting the arguments and outcomes of the update rule $\phi_P$). In this case, the resultant fixed-point outcomes of $V$ are already ensured to be  inverted. 
We again point out the analogy with the ``inside'' self-reference in formal systems visible here in the CA $M$ running on an encoding of itself. 

Finally, we consider dynamics of the inverter  $V$ running with the initial configuration $v^0 = [V]$ (this is, of course, similar to the construction of the ``external'' self-reference in formal systems), which corresponds to the following constraint, resulting from substituting the elements of $V$ for the elements of $M$ in expression (\ref{eqCA:V}):
\begin{equation}
\label{ca-vv}
\begin{cases}
V: [V] \rightarrow v^-  \hspace{7mm} \textrm{whenever} \ V: [V] \rightarrow v^+    \\%[0.8em]
V: [V] \rightarrow v^+  \hspace{7mm} \textrm{whenever} \ V: [V] \rightarrow v^- \  \textrm{or runs forever} 
\end{cases}
\end{equation}
The result is again a contradiction in the style of the Liar's Paradox: the CA $V$ reaches an  attractor in the subset $\Psi_V^-$ whenever it reaches  an attractor in the complement subset $\Psi_V^+$. 
%On the other hand, CA $V$ reaches a fixed-point in the subset $\Psi_V^+$ whenever it reaches a fixed-point in the complement subset $\Psi_V^-$ (as CA $V$ cannot run forever, being an inverter of the decider $P$).
This contradiction shows the impossibility of the existence of a decider CA $P$, and therefore, the undecidability of CA dynamics. 
We note that the inverter CA $V$ was running on the input representing its own description $[V]$, while employing the  decider CA $P$ ``sandwiched'' between the self-referencing $V$ and the self-referencing $M$. 

Continuing with the ``Life in Life'' $V$ example, we can express this contradiction through the meta-level dynamics reaching the configuration that corresponds to the ``accepted'' outcome, being in $\Psi_V^+$, but at the underlying level of the CA $V$ itself this configuration indicates the ``rejected'' outcome, being in $\Psi_V^-$. This forms a contradiction only because the CA $V$ simulates itself. We must note that a key step leading to the contradiction is the inversion of the termination condition which occurred \emph{outside} of the system \emph{per se}. Thus, it can be argued that this contradiction is empowered not only by the ability to represent programs as data (via suitable encodings) and the ability to design universal CAs, but also by the capacity to assign a negative meaning to the observed configurations. This is, in fact, the same mechanism that was employed in  G\"odel's proof where the Self-reference lemma was applied to the \emph{negated} provability predicate $\neg \textrm{Provable}_{\altmathcal{F}}(x)$.

%One could implement the Liar's Paradox within a dynamical system in other ways, so that the paradox takes the form  ``the system is not stable if and only if it can be shown to be stable'', as discussed by  \cite{Hyotyniemi}.

We re-iterate that universal CAs are definitely constructable and as we pointed out, the CA rule $110$ and ``Life in Life'' have been shown to be capable of universal computation \cite{Cook2004,Cook2004,wiki:octa}. What is actually impossible %, in order to preserve the undecidable dynamics, 
is a specification of a definitive termination condition  assigning binary outcomes for any possible CA $M$ that is being simulated, as in (\ref{ca-h}).

\section{Results}

\subsection{Diagonalization and self-reference}
\label{diag}

To illustrate the diagonalization argument employed in the undecidability proof(s) in various  frameworks, we follow the  expositions offered by Buldt \cite{Buldt2016} and Gaifman \cite{Gaifman2007easy} in the context of formal systems, adapted for our purposes.

In Step 1, the (at most countable) set of all  first-order expressions with the free variable $x$ is considered:
\[
\altmathcal{A} = \{W_0(x), W_1(x), W_2(x), \ldots \}.
\]

In Step 2, the set of all of their G\"odel numbers is formed:
\[
\altmathcal{B} = \{\ulcorner \ W_0(x) \ \urcorner, \ \ulcorner \ W_1(x) \ \urcorner, \ \ulcorner \ W_2(x) \ \urcorner, \ \ldots \}.
\]

In Step 3,  all members of set $\altmathcal{B}$ are used in place of the free variables of all members of the set $\altmathcal{A}$.  Denoting $W_{ij} = W_i(\ulcorner \ W_j(x) \ \urcorner)$, a matrix is constructed as shown in Table \ref{Decider_Table_FS}.

\begin{table}[h]
\begin{center}
\begin{tabular}{c|cccc}
& $\ulcorner \ W_0(x) \ \urcorner$ & $\ulcorner \ W_1(x) \ \urcorner$ & $\ulcorner \ W_2(x) \ \urcorner$ & $\cdots$   \\
\hline
$W_0(x)$ & $W_{00}$ & $W_{01}$ & $W_{02}$ &     
\\
$W_1(x)$ & $W_{10}$ & $W_{11}$ & $W_{12}$ & $\cdots$    
\\
$W_2(x)$ & $W_{20}$ & $W_{21}$ & $W_{22}$ &    
\\
$\vdots$ &  
& $\vdots$ &
&$\ddots$     \\
\end{tabular}
\end{center}
\caption{\label{Decider_Table_FS} First diagonalization (i.e., ``internal'' self-reference) for a formal system.}
\end{table}

The diagonal sequence $\{W_{jj}\}$ corresponds to the ``first diagonalization'' (i.e., first, or ``internal'', self-reference \cite{Buldt2016,Gaifman2007easy}).  

%Noting that the predicate $\neg \textrm{Provable}_{\altmathcal{F}}(x)$ is itself a member of the set $\altmathcal{A}$, there must be an index $k$ such that $W_k(x) \equiv \neg \textrm{Provable}_{\altmathcal{F}}(x)$, and so the entities filling the $k$'s row of the table are $W_{kj} = \neg \textrm{Provable}_{\altmathcal{F}}(\ulcorner \ W_j(x) \ \urcorner)$. 

The next step is to consider the row of the table, with an index $k$, corresponding to the predicate 
\[W_k(x) \equiv \neg \textrm{Provable}_{\altmathcal{F}}(\textrm{diag}(x) ) \ ,
\]
 where the term $\textrm{diag}(x)$ corresponds to a function $diag(x)$ that maps the G\"odel number of a wff \ $W(x)$ to the G\"odel number of the self-referential wff \ $W(\ulcorner \ W(x) \ \urcorner)$, that is: 
\[diag(\altmathcal{G}(W(x))) \equiv  \altmathcal{G}(W(\ulcorner \ W(x) \ \urcorner))   
\] 
and
\[\textrm{diag}(\ulcorner \ W(x) \ \urcorner) =  \ulcorner \ W(\ulcorner \ W(x) \ \urcorner) \ \urcorner   \ .
\] 
As pointed out by Gaifman, it does not matter how the function $diag(x)$ is defined on numbers that are not G\"odel numbers \cite{Gaifman2007easy}. 
%Specifically, we can obtain a formal expression $\textrm{diag}(\ulcorner \ W_j(x) \ \urcorner) = \ulcorner \ W_j(\ulcorner \ W_j(x) \ \urcorner) \ \urcorner$ for any element in the set $\altmathcal{B}$.
The elements of the $k$'th row are formed, as any other elements of the table, by using all members of set $\altmathcal{B}$, i.e., the numerals $\ulcorner \ W_j(x) \ \urcorner$, in place of the free variable of the  predicate $W_k(x)$:
\[W_{kj} = \neg \textrm{Provable}_{\altmathcal{F}}(\ulcorner \ W_j(\ulcorner \ W_j(x) \ \urcorner) \ \urcorner) \ .\] 
 In the style of Cantor's diagonalization method, we can informally say that the  $k$'th row of the table  ``inverts'' the diagonal entities $W_{jj} = W_j(\ulcorner \ W_j(x) \ \urcorner)$, by applying $\neg \textrm{Provable}_{\altmathcal{F}}$ to numerals of their G\"odel numbers.  
Importantly, the predicate $W_k(x) \equiv \neg \textrm{Provable}_{\altmathcal{F}}(\textrm{diag}(x) )$ is itself a member of the set $\altmathcal{A}$,  by construction being distinct from other members $W_j(x)$, see Table \ref{Decider_Table_FS2}.

\begin{table}[h]
\begin{center}
\begin{tabular}{c|cccccc}
& $\ulcorner \ W_0(x) \ \urcorner$ & $\ulcorner \ W_1(x) \ \urcorner$ & $\ulcorner \ W_2(x) \ \urcorner$ & $\cdots$ & $\ulcorner \ W_k(x) \ \urcorner$ & $\cdots$  \\
\hline
$W_0(x)$ & $W_{00}$ & $W_{01}$ & $W_{02}$ & & $W_{0k}$ &    
\\
$W_1(x)$ & $W_{10}$ & $W_{11}$ & $W_{12}$ & $\cdots$& $W_{1k}$ & $\cdots$    
\\
$W_2(x)$ & $W_{20}$ & $W_{21}$ & $W_{22}$ & & $W_{2k}$ &    
\\
$\vdots$ &  
& $\vdots$ &
&$\ddots$ &  
&     \\
$W_k(x)$ & $W_{k0}$ & $W_{k1}$ & $W_{k2}$ & & $W_{kk} = \gamma$
&     \\
$\vdots$ &  
& $\vdots$ &
& &  
&   $\ddots$
\end{tabular}
\end{center}
\caption{\label{Decider_Table_FS2} Second diagonalization (i.e., ``external'' self-reference) for a formal system.}
\end{table}

The crux of the argument is the element $W_{kk} = \neg \textrm{Provable}_{\altmathcal{F}}(\ulcorner \ W_k(\ulcorner \ W_k(x) \ \urcorner) \ \urcorner)$ which was also technically formed, at Step 3 above, as $W_{kk} = W_k(\ulcorner \ W_k(x) \ \urcorner)$. Finally we arrive at a  G\"odel sentence $\gamma = W_k(\ulcorner \ W_k(x) \ \urcorner)$ which is neither provable nor disprovable in $\altmathcal{F}$, cf. key expressions (\ref{negProv}) and (\ref{vv}) re-expressed in terms of $\gamma$:
\begin{equation}
\label{gamma}
  \altmathcal{F} \vdash \ \gamma \leftrightarrow \neg \textrm{Provable}_{\altmathcal{F}}(\ulcorner \ \gamma \ \urcorner) \ . 
\end{equation}
Again, in forming the diagonal element $W_{kk}$, the G\"odel sentence $\gamma$ is self-referencing:  this is the second diagonalization \cite{Buldt2016} or second, ``external'' use of self-reference \cite{Gaifman2007easy}.

The diagonalization argument presented above can be seen almost as a template, including the first diagonalization  $W_{ij} = W_i(\ulcorner \ W_j(x) \ \urcorner)$, then the ``inversion'' $\neg \textrm{Provable}_{\altmathcal{F}}$ applied to numerals of G\"odel numbers of the diagonal elements, and finally the second diagonalization where we construct the G\"odel sentence $\gamma = W_k(\ulcorner \ W_k(x) \ \urcorner)$ used in expression (\ref{gamma}).

Using this template we can now apply the  diagonalization argument to show the undecidability of both TMs,  (\ref{t:p})--(\ref{t:dd}), and CAs, (\ref{ca-h})--(\ref{ca-vv}).

In the Table \ref{Decider_Table1} the rows correspond to all TMs (or CAs) $M_1, M_2, \ldots, M_j, \ldots$, and the columns correspond to the encodings of these objects $[M_1], [M_2], \ldots, [M_j], \ldots$. Each element of the table is `accept' if the machine accepts the input but is blank if it rejects or loops on that input, cf. expression (\ref{t:p}) \cite{Sipser}. In case of CAs, `accept' represents the outcome $M_i: [M_j] \rightarrow m^+$, and blank then represents the outcome $M_i: [M_j] \rightarrow m^- \  \textrm{or runs forever}$, corresponding to expression (\ref{ca-h}).

\begin{table}[h]
\begin{center}
\begin{tabular}{c|ccccc}
& $[ M_1 ]$ & $[ M_2 ]$ & $[ M_3 ]$ & $\cdots$   \\
\hline
$M_1$ & accept & & accept &
\\
$M_2$ & accept & accept & accept & $\cdots$& 
\\
$M_3$ &  & accept &  &
\\
$\vdots$ &  
& $\vdots$ &
&$\ddots$      \\
\end{tabular}
\end{center}
\caption{\label{Decider_Table1} The cell $i,j$ is `accept' if $M_i$ accepts $[M_j]$, (or for the CAs: $M_i: [M_j] \rightarrow m^+$).}
\end{table}

The assumption that there exists a decider TM $P$ (or decider CA $P$) corresponds to ``filling'' the table with `reject' entries in place of the blanks, as every program-data combination is assumed to be decidable, shown in Table \ref{Decider_Table2} \cite{Sipser}. For example, if $M_3$ does not accept the input $[M_1]$, the entry $(3,1)$ is now `reject' because the decider machine or decider CA $P$ rejects the input $[M_3[M_1]]$, cf. expressions (\ref{t:p}) and (\ref{ca-h}).

\begin{table}[h]
\begin{center}
\begin{tabular}{c|ccccc}
& $[ M_1 ]$ & $[ M_2 ]$ & $[ M_3 ]$ & $\cdots$   \\
\hline
$M_1$ & accept & reject & accept &
\\
$M_2$ & accept & accept & accept & $\cdots$& 
\\
$M_3$ & reject & accept & reject &
\\
$\vdots$ &  
& $\vdots$ &
&$\ddots$      \\
\end{tabular}
\end{center}
\caption{\label{Decider_Table2} The cell $i,j$ is the outcome of running $P$ on $[M_i[M_j]]$.}
\end{table}

The diagonal sequence is a result of the first diagonalization (first self-reference), and we  can now invert the diagonal elements in order to populate the row representing the inverter TM (CA) $V$, analogously to the construction of Table \ref{Decider_Table_FS2} for formal systems, and matching the expressions (\ref{t:d}) and (\ref{eqCA:V}).  The  result of including the inverter machine (CA) $V = M_k$, for some $k$, is shown in Table \ref{Decider_Table}, where the element $(k,k)$ is an analogue of the G\"odel sentence $\gamma = W_k(\ulcorner \ W_k(x) \ \urcorner)$: the inverter machine (CA) $V = M_k$ runs on $[M_k[M_k]]$, which is the second diagonalization. Neither `accept' nor `reject' in place of the element $(k,k)$ would avoid a logical contradiction. This refutes the assumption of the existence of the decider machine (CA) $P$.

\begin{table}[h]
\begin{center}
\begin{tabular}{c|cccccc}
& $[ M_1 ]$ & $[ M_2 ]$ & $[ M_3 ]$ & $\cdots$ &$[ M_k ]$ & $\cdots$  \\
\hline
$M_1$ & \underline{accept} & reject
& accept &
& accept &    
\\
$M_2$ & accept
& \underline{accept}
& accept & $\cdots$& reject & $\cdots$    
\\
$M_3$ & reject
& accept & \underline{reject} &
& reject &    
\\
$\vdots$ &  
& $\vdots$ &
&$\ddots$ &  
&     \\
$M_k$ & reject
& reject & accept &
& \underline{?}
&     \\
$\vdots$ &  
& $\vdots$ &
& &  
&   $\ddots$
\end{tabular}
\end{center}
\caption{\label{Decider_Table} The cell $(k,j)$ is the outcome of running $V = M_k$ (the inverter of $P$) on $[M_k[M_j]]$. A contradiction occurs at cell $(k,k)$.}
\end{table}

%This decider machine (CA) $P$  was used in ``filling'' the blanks of Table \ref{Decider_Table1}. Contrasting with Table \ref{Decider_Table_FS} for formal systems, which was ``pre-filled'', we note that the role of decider $P$ was played by the representative predicate $\textrm{P}$.  

%\vspace*{-2mm}
One instructive comparison is that the decoding and encoding sub-steps used in creating $[M_k[M_k]]$ are analogous to the 
 function $diag(x)$ that maps the G\"odel number of a formula \ $W_k(x)$ to the G\"odel number of the self-referential formula  $\ulcorner \ W_k(\ulcorner \ W_k(x) \ \urcorner) \ \urcorner$. That is, given $[M_k]$ or $\ulcorner \ W_k(x) \ \urcorner$ one may choose to decode into   $M_k$ or $W_k(x)$, and then run the decoded machine (CA) or use the formula $W_k(x)$ on itself, constructing the final self-referential input $[M_k[M_k]]$ or $\ulcorner \ W_k(\ulcorner \ W_k(x) \ \urcorner) \ \urcorner$.

\subsection{Comparative Analysis}
\label{sum}

We have considered three computational frameworks (formal systems, Turing machines and Cellular Automata), focussing on self-reference, diagonalization and undecidability manifested on a fundamental level. In this section we offer a  detailed comparative analysis across specific structural elements utilized in these frameworks. In doing so, we separately analyze different ways to structure the state-space, define the problem, and evolve the system's dynamics, culminating with a comparison of the mechanics of undecidability. 
While some of these comparisons are well-noted in the literature at a high level \cite{Cas91,Ali1999,Ilachinski2001}, the rest, we believe, reveals the deeper formal analogies unifying the frameworks at a much more detailed level.  

\subsubsection{State-space}

The three computational frameworks that we considered define their state-space in different but analogous terms, and Table \ref{Summary} explicitly contrasts the corresponding formal descriptions.

\begin{table}[h]
\begin{center}
\renewcommand{\arraystretch}{1.5}
\begin{tabular}{|p{4.2cm}|p{5cm}|p{5.8cm}|}
\hline
 Formal systems & Turing machines & Cellular Automata  \\ 
\hline \hline
alphabet $\altmathcal{A}_{\altmathcal{F}}$ & alphabets: input $\Sigma$ and tape $\Gamma$ &  alphabet $A_C$  \\
\hline
symbol strings in $\altmathcal{A}^*_{\altmathcal{F}}$ & tape strings in $\Sigma^*$  & configurations in state-space $\Psi_C = A_C^{\mathbb{Z}^d}$  \\
\hline
grammar $\langle \altmathcal{A}_{\altmathcal{F}}, \altmathcal{N}_{\altmathcal{F}}, \altmathcal{P}_{\altmathcal{F}}, \altmathcal{S}_{\altmathcal{F}} \rangle$ & admissible syntax, given $\Gamma$, e.g., blank symbol & constraints on state-space $\Psi_C$, e.g., by recursive configurations \\
\hline
well-formed formula in $\altmathcal{A}^*_{\altmathcal{F}}$, restricted by grammar & recognizable tape pattern in $\Sigma^*$, given $\Gamma$ & primitive recursively decidable configuration in restricted subset of $\Psi_C$ \\
\hline
infinite language & infinite tape & infinite lattice \\ 
\hline
\end{tabular}
\end{center}
\caption{\label{Summary} State-space comparison across three computational frameworks.}
\end{table}

The three row elements  describing grammar/syntax/restriction must ensure that the well-formed formulas, tape patterns and CA configurations are effectively computable. In formal systems this guarantees that a decision procedure for deciding whether a formula is
well-formed or not does exist;  the tape patterns of a TM are recognizable; and in CAs  an ``effective dynamical system'' is maintained.

\subsubsection{Problem definition and dynamics}

The adopted definition of a TM used two final  states $q_{acc}$ and $q_{rej}$ to distinguish whether the initial input (which includes a target problem to be solved) is accepted or rejected.  To re-iterate, according to this definition, denoted ($\ddagger$), the initial tape input includes the target to be verified (to be either accepted or rejected), and therefore, the final  content of the tape, upon halting at either $q_{acc}$ or $q_{rej}$, does not matter. As mentioned, an equivalent  definition of a TM, denoted ($\dagger$), may have just one halting state $q_{halt}$, in which case the target is not included on the tape's initial input, but when the machine halts, the content written on the tape represents the actual output of the computation. 

The elements describing axioms and initial inputs/configurations, shown in Table \ref{Summary2}, leave some room in the initial conditions to also include the target statement: a theorem (in a formal system), a target to be verified (by a TM), or a target configuration (of a CA extended with a termination condition).

\begin{table}[h]
\begin{center}
\renewcommand{\arraystretch}{1.5}
\begin{tabular}{|p{5.3cm}|p{5cm}|p{4.7cm}|}
\hline
 Formal systems & Turing machines & Cellular Automata  \\ 
\hline \hline
axioms $X_{\altmathcal{F}}$ & (a part of) initial tape & (a part of) initial configuration \\
\hline
($\ddagger$) target: well-formed formula (wff) to be proven & ($\ddagger$) target: \newline string as part of initial tape & ($\ddagger$) target: \newline subset of initial configuration \\
\hline
($\ddagger$) proving or disproving a target wff & ($\ddagger$) final states $q_{acc}$ and $q_{rej}$ & ($\ddagger$) termination condition testing against $\Psi_C^+$ or $\Psi_C^-$ \\
\hline
rules of inference $R_{\altmathcal{F}}$ & transition function $\mu$ &  local update rule $\phi_C$ \\
\hline
proof: derivation sequence & sequence of tape patterns and machine states & dynamics: evolution of configurations \\
\hline
($\dagger$) an external criterion distinguishing a wff in a proof & ($\dagger$) final state $q_{halt}$ & ($\dagger$) termination condition testing for fixed points or limit cycles \\
\hline
($\dagger$) theorem: the last wff in a proof & ($\dagger$) final output written on the tape & ($\dagger$) the attractor configuration(s) \\
\hline
\end{tabular}
\end{center}
\caption{\label{Summary2} Problem definition and inferences/computation/dynamics in three computational frameworks.}
\end{table}

Having considered the elements that define the problem and drive the system's ``evolution'', that is, the inference process within a formal system, the computation by a TM, or the CA dynamics, we now turn our attention to the mechanics employed by the different proofs of undecidability in our three computational frameworks.

\subsubsection{Undecidable dynamics}

Table \ref{Summary3} traces the key steps of the diagonalization argument. The existence of some elements in the table have only been assumed for the purposes of proof by contradiction, and we denote these lines by $\nexists$.
%Importantly,  G\"odel's First Incompleteness Theorem centered on G\"odel sentences $V^{\textrm{P}}$, relative to representative predicate $\textrm{P}(x)$, can be used to demonstrate undecidability \cite{Raatikainen}.  In other words, the G\"odel sentence $\gamma$ can be expressed but not proven within a formal system 

\begin{table}[h]
\begin{center}
\renewcommand{\arraystretch}{1.8}
\begin{tabular}{|p{5.2cm}|p{5cm}|p{4.8cm}|}
\hline
 Formal systems & Turing machines & Cellular Automata  \\ 
\hline \hline
weakly representative predicate \newline (``a mocking bird'') & universal TM & universal CA \\
\hline 
($\nexists$)  representative predicate $\textrm{P}(x)$ & ($\nexists$)  decider UTM $P$ & ($\nexists$)  universal decider CA $P$ \\
\hline
G\"odel number of $W_j(x)$, denoted  $\altmathcal{G}(W(x))$, such that $\ulcorner \ W \ \urcorner = \underline{\altmathcal{G}(W)}$  & encoding of TM $M_j$, denoted $[M_j]$ & encoding of CA $M_j$, denoted  $[M_j]$ \\
\hline
unique decoding of $W(x)$ from G\"odel number $\altmathcal{G}(W(x))$ & unique decoding of TM $M$ from $[M]$ & unique decoding of CA $M$ from $[M]$ \\
\hline
first diagonalization, internal self-referencing: $W_j(\ulcorner \ W_j(x) \ \urcorner)$ & first diagonalization, ``internal'' self-referencing: $M_j[M_j]$ &  first diagonalization, ``internal'' self-referencing: $M_j[M_j]$ \\
\hline
diagonalization term for $W(x)$: $\textrm{diag}(\ulcorner  W(x)  \urcorner) = \ulcorner  W(\ulcorner  W(x)  \urcorner)  \urcorner$ 
& compound encoding of TM $M$, as $[M[M]]$ & compound encoding of CA $M$, as $[M[M]]$ \\ 
\hline
``inverted'' predicate \newline $V^{\textrm{P}}(x) \equiv \neg \textrm{P}_{\altmathcal{F}}(\textrm{diag}(x) )$ & inverter TM $V([M])$ & inverter CA $V$ running on $[M]$ \\ 
\hline
G\"odel sentence \newline $V^{\textrm{P}}(x) = V(\ulcorner \ V(x) \urcorner)$ & self-referencing inverter TM $V([V])$  & inverter CA $V$ running on $[V]$ \\
\hline
second diagonalization, external self-referencing: $\altmathcal{F} \xvdash{?}  \ V^{\textrm{P}} \leftrightarrow \neg \textrm{P}(\ulcorner \ V^{\textrm{P}} \ \urcorner)$ & second diagonalization, external self-referencing: $V([V]) = \ ?$ & second diagonalization, external self-referencing: $V: [V] \rightarrow v^?$ \\ 
\hline
\hline
G\"odel Incompleteness Theorem, leading to undecidability & The Halting Problem & Undecidable dynamics and the ``Edge of Chaos'' \\
\hline
\end{tabular}
\end{center}
\caption{\label{Summary3} Proving undecidability in three computational frameworks.}
\end{table}

Importantly, each of the proofs ends up in a contradiction. In formal systems, the proof constructs a G\"odel sentence which yields a contradiction, expressed as an inability to resolve the question: $\altmathcal{F} \xvdash{?}  \ V^{\textrm{P}} \leftrightarrow \neg \textrm{P}(\ulcorner \ V^{\textrm{P}} \ \urcorner)$. This results in the G\"odel Incompleteness Theorem that leads to undecidability. In TMs, the contradiction comes from trying to answer the halting question about the inverter machine running on the encoding of itself: $V([V]) = \ ?$ which is the core issue of The Halting Problem. And in Cellular Automata, the conundrum manifests itself as the question of whether the inverter CA would reach a termination condition if presented with an initial condition that encodes its own description, $V: [V] \rightarrow v^?$.  This, in our opinion, captures undecidable dynamics at the ``edge of chaos''. 

\section{Discussion}

It is important to point out that in all considered computational frameworks the undecidable ``dynamics'' are possible even with perfect  knowledge of the initial / boundary conditions of the system \cite{Bennett1990,Moore1990,Moore1991}. As mentioned in Introduction, this distinguishes undecidable dynamics from chaotic dynamics.  Interestingly, undecidable dynamics can also be distinguished from  unprestatable functions and their  evolution \cite{Kauffman2016}: when a dynamical system (e.g., a chemical reaction system) alters its own boundary condition,  we cannot deduce the actual behavior of the system even from the same initial and boundary conditions. Unprestatable dynamics resulting from such, possibly iterative, modifications of the boundary conditions is also unlike standard chaos. However, the class of systems with dynamically altering boundaries and hence, unprestatable dynamics, is distinct from the systems with undecidable dynamics which evolve from fixed  initial conditions.  

Therefore, one may be justified in defining a complex system as a dynamical system with at least undecidable dynamics, and possibly unprestatable dynamics.

As has been previously pointed out \cite{casti94,Markose2004b,prok2014,Markose2017}, undecidability may be fundamentally related to  computational novelty, and so a mechanism producing  novelty may need to be capable of universal computation. For example, Markose \cite{Markose2017} recently argued that the issue of novelty production and ``thinking outside the box'' by digital agents must be immediately related to their capacity to encode a G\"odel sentence in order to exit from known listable sets (e.g., actions, technologies, phenotypes) and produce new structured objects.  This formalism follows Binmore \cite{binmore1987} in highlighting the fundamental aspects of novelty generation through the lens of game-theory, and considering a strategic game with adversarial (contrarian) agents which act as the Liar by negating what it can predict or compute \cite{Markose2017}. It has also been recently argued that evolutionary strategies in iterated games, in which the same economic interaction is repeatedly played between the same agents, can be seen as processes capable of universal computation \cite{harre2017}. In other words, undecidable dynamics is the necessity for creativity and innovation.

As we have shown, the capacity to generate undecidable dynamics is based upon three underlying factors: (i) the program-data duality; (ii) the potential to access an infinite computational medium; and (iii) the ability to implement negation.  It is interesting to note parallels between these principles and Markose's ingredients for novelty generation by digital agents, underpinned by G\"odel -- Turing -- Post approach \cite{Markose2017}: (1) agents can operate on encoded information and store codes; (2) agents can do offline simulations that involve self-referential meta-calculations, i.e., deal with G\"odel meta-mathematics; and (3)  agents can record negation and, therefore, ``can process the logical archetype of the Liar in a fixed point setting''.  These considerations emphasize once more the self-referential basis of undecidable dynamics, not only providing foundations for the most general computational frameworks, but also revealing paths for implementing   complex adaptive systems.

%\section*{Acknowledgements}

\bibliographystyle{elsarticle-num}
%\bibliography{TE}

\begin{thebibliography}{10}
\expandafter\ifx\csname url\endcsname\relax
  \def\url#1{\texttt{#1}}\fi
\expandafter\ifx\csname urlprefix\endcsname\relax\def\urlprefix{URL }\fi
\expandafter\ifx\csname href\endcsname\relax
  \def\href#1#2{#2} \def\path#1{#1}\fi

\bibitem{Cas91}
J.~L. Casti, {Chaos, G\"{o}del and Truth}, in: J.~L. Casti, A.~Karlqvist
  (Eds.), Beyond Belief: Randomness, Prediction, and Explanation in Science,
  CRC Press, 1991.

\bibitem{Ilachinski2001}
A.~Ilachinski, Cellular Automata: A Discrete Universe, World Scientific,
  Singapore, 2001.

\bibitem{Bennett1990}
C.~H. Bennett, Undecidable dynamics, Nature 346 (1990) 606--607.

\bibitem{Moore1990}
C.~Moore, Unpredictability and undecidability in dynamical systems, Physical
  Review Letters 64~(20) (1990) 2354--2357.

\bibitem{Moore1991}
C.~Moore, Generalized shifts: unpredictability and undecidability in dynamical
  systems, Nonlinearity 4~(2) (1991) 199--230.

\bibitem{Durand}
B.~Durand, E.~Formenti, G.~Varouchas, On undecidability of equicontinuity
  classification for cellular automata, in: {DMCS}, Vol.~{AB} of Discrete
  Mathematics and Theoretical Computer Science Proceedings, {DMTCS}, 2003, pp.
  117--128.

\bibitem{Kari}
J.~Kari, Decidability and undecidability in cellular automata, International
  Journal of General Systems 41~(6) (2012) 539--554.

\bibitem{Sutner}
K.~Sutner, Computational classification of cellular automata, International
  Journal of General Systems 41~(6) (2012) 595--607.

\bibitem{Wolfram1984}
S.~Wolfram, Computation theory of cellular automata, Communications in
  Mathematical Physics 96~(1) (1984) 15--57.

\bibitem{wolfram1985}
S.~Wolfram, Twenty problems in the theory of {Cellular Automata}, Physica
  Scripta 1985~(T9) (1985) 170.

\bibitem{Wolfram2002}
S.~Wolfram, A New Kind of Science, Wolfram Media Inc., Champaign, Ilinois, US,
  United States, 2002.

\bibitem{lang90}
C.~G. Langton, Computation at the edge of chaos: phase transitions and emergent
  computation, Physica D 42~(1-3) (1990) 12--37.

\bibitem{crutch94}
J.~P. Crutchfield, The calculi of emergence: computation, dynamics and
  induction, Physica D 75~(1-3) (1994) 11--54.

\bibitem{wue99}
A.~Wuensche, Classifying cellular automata automatically: Finding gliders,
  filtering, and relating space-time patterns, attractor basins, and the {Z}
  parameter, Complexity 4~(3) (1999) 47--66.

\bibitem{hord01}
W.~Hordijk, C.~R. Shalizi, J.~P. Crutchfield, Upper bound on the products of
  particle interactions in cellular automata, Physica D 154~(3-4) (2001)
  240--258.

\bibitem{sha06}
C.~R. Shalizi, R.~Haslinger, J.-B. Rouquier, K.~L. Klinkner, C.~Moore,
  Automatic filters for the detection of coherent structure in spatiotemporal
  systems, Physical Review E 73~(3) (2006) 036104.

\bibitem{liz08a}
J.~T. Lizier, M.~Prokopenko, A.~Y. Zomaya, Local information transfer as a
  spatiotemporal filter for complex systems, Physical Review E 77~(2) (2008)
  026110.

\bibitem{liz12a}
J.~T. Lizier, M.~Prokopenko, A.~Y. Zomaya, Local measures of information
  storage in complex distributed computation, Information Sciences 208 (2012)
  39--54.
\newblock \href {http://dx.doi.org/10.1016/j.ins.2012.04.016}
  {\path{doi:10.1016/j.ins.2012.04.016}}.

\bibitem{pack88}
N.~H. Packard, Adaptation toward the edge of chaos, in: J.~A.~S. Kelso, A.~J.
  Mandell, M.~F. Shlesinger (Eds.), Dynamic Patterns in Complex Systems, World
  Scientific, 1988, pp. 293--301.

\bibitem{Crutchfield1988}
J.~P. Crutchfield, K.~Young, Computation at the onset of chaos, in: The Santa
  Fe Institute, Westview, Press, 1988, pp. 223--269.

\bibitem{mitch94b}
M.~Mitchell, J.~P. Crutchfield, P.~T. Hraber, Dynamics, computation, and the
  "edge of chaos": A re-examination, in: G.~Cowan, D.~Pines, D.~Melzner (Eds.),
  Complexity: Metaphors, Models, and Reality, Vol.~19 of Santa Fe Institute
  Studies in the Sciences of Complexity, Addison-Wesley, Reading, MA, 1994, pp.
  497--513.

\bibitem{liz08c}
J.~T. Lizier, M.~Prokopenko, A.~Y. Zomaya, The information dynamics of phase
  transitions in random {B}oolean networks, in: S.~Bullock, J.~Noble,
  R.~Watson, M.~A. Bedau (Eds.), Proceedings of the Eleventh International
  Conference on the Simulation and Synthesis of Living Systems (ALife XI),
  Winchester, UK, MIT Press, Cambridge, MA, 2008, pp. 374--381.

\bibitem{liz11b}
J.~T. Lizier, S.~Pritam, M.~Prokopenko, Information dynamics in small-world
  {B}oolean networks, Artificial Life 17~(4) (2011) 293--314.
\newblock \href {http://dx.doi.org/10.1162/artl\_a\_00040}
  {\path{doi:10.1162/artl\_a\_00040}}.

\bibitem{boed12a}
J.~Boedecker, O.~Obst, J.~T. Lizier, Mayer, M.~Asada, Information processing in
  echo state networks at the edge of chaos, Theory in Biosciences 131~(3)
  (2012) 205--213.

\bibitem{Siegelmann1999}
H.~T. Siegelmann, Neural Networks and Analog Computation: Beyond the Turing
  Limit, Birkhauser Boston Inc., Cambridge, MA, USA, 1999.

\bibitem{delvenne2006decidability}
J.-C. Delvenne, P.~K\r{u}rka, V.~Blondel, Decidability and universality in
  symbolic dynamical systems, Fundamenta Informaticae 74~(4) (2006) 463--490.

\bibitem{Zenil2013}
G.~J. Mart{\'{\i}}nez, J.~C. S.~T. Mora, H.~Zenil, {Computation and
  Universality: Class {IV} versus Class {III} Cellular Automata}, Journal of
  Cellular Automata 7~(5-6) (2012) 393--430.

\bibitem{Sutner-chapter}
K.~Sutner, Cellular automata , classification of, in: R.~A. Meyers (Ed.),
  Computational Complexity: Theory, Techniques, and Applications, Artificial
  Intelligence, Springer, 2012, pp. 312--324.

\bibitem{Hyotyniemi}
H.~Hy\"otyniemi, On the universality and undecidability in dynamic systems,
  Technical Report 133, Control Engineering Laboratory, Helsinki University of
  Technology (2002).

\bibitem{Ali1999}
S.~M. Ali, The concept of poi{\=e}sis and its application in a heideggerian
  critique of computationally emergent artificiality, Ph.D. thesis, Department
  of Electrical \& Electronic Engineering, Brunel University (1999).

\bibitem{Smullyan1961}
R.~M. Smullyan, Theory of formal systems, Princeton University Press, 1961.

\bibitem{Buldt2016}
B.~Buldt, On fixed points, diagonalization, and self-reference, in: W.~F.
  et~al. (Ed.), Von Rang und Namen. Essays in Honour of Wolfgang Spohn,
  M\"unster, Mentis, 2016, pp. 47--63.

\bibitem{Rapaport2005}
W.~J. Rapaport, Philosophy of computer science: An introductory course,
  Teaching Philosophy 28~(4) (2005) 319--341.

\bibitem{Chomsky1956}
N.~Chomsky, Three models for the description of language, IRE Transactions on
  information theory 2~(3) (1956) 113--124.

\bibitem{Nagel2001}
E.~Nagel, J.~R. Newman, G{\"o}del's proof, New York University Press, New York,
  2001.

\bibitem{Raatikainen}
P.~Raatikainen, G\"{o}del's incompleteness theorems, in: E.~N. Zalta (Ed.), The
  Stanford Encyclopedia of Philosophy, spring 2015 Edition, Metaphysics
  Research Lab, Stanford University, 2015.

\bibitem{Gaifman2006naming}
H.~Gaifman, Naming and diagonalization, from {Cantor to G{\"o}del to Kleene},
  Logic Journal of the IGPL 14~(5) (2006) 709--728.

\bibitem{Smullyan1984}
R.~M. Smullyan, Fixed points and self-reference, International Journal of
  Mathematics and Mathematical Sciences 7~(2) (1984) 283--289.

\bibitem{Carnap1934}
R.~Carnap, {Logische Syntax der Sprache. Schriften zur wissenschaftlichen
  Weltauffassung (1934). In English: The Logical Syntax of Language}, Routledge
  and Kegan Paul, London, 1937, 1971 printing.

\bibitem{Tarski1936}
A.~Tarski, {Der Wahrheitsbegriff in den formalisierten Sprachen (1936). In
  English: The Concept of Truth in Formal Systems}, in: Logic, semantics,
  metamathematics: papers from 1923 to 1938 / by A. Tarski. Translated from
  various languages by J.H.Woodger, Clarendon Press Oxford, 1956.

\bibitem{Sieg2005}
W.~Sieg, C.~Field, Automated search for {G{\"{o}}del's} proofs, Annals of Pure
  and Applied Logic 133~(1-3) (2005) 319--338.

\bibitem{godel2003collected}
K.~G{\"o}del, {\"Uber formal unentscheidbare S\"atze der Principia Mathematica
  und verwandter Systeme, I (1931)}, in: S.~Feferman (Ed.), Collected works.
  Vol. 1, Publications 1929-1936, Oxford University Press, 1986.

\bibitem{Gaifman2007easy}
H.~Gaifman, The easy way to {G{\"o}del}'s proof and related matters,
  \url{http://www.columbia.edu/~hg17/Inc07-chap0.pdf} (2007).

\bibitem{Sipser}
M.~Sipser, Introduction to the Theory of Computation, 1st Edition,
  International Thomson Publishing, 1996.

\bibitem{Hopcroft1969}
J.~E. Hopcroft, J.~D. Ullman, Formal languages and their relation to automata,
  Addison-Wesley Publishing Company, Reading, Massachusetts, 1969.

\bibitem{goldenfeld}
N.~Israeli, N.~Goldenfeld, Coarse-graining of cellular automata, emergence, and
  the predictability of complex systems, Physical Review E 73~(2).

\bibitem{Cook2004}
M.~Cook, Universality in elementary cellular automata, Complex Systems 15~(1)
  (2004) 1--40.

\bibitem{Gardner1970}
M.~Gardner, {Mathematical Games: The fantastic combinations of John Conway's
  new solitaire game "life"}, Scientific American 223 (1970) 120--123.

\bibitem{Berlekamp1982}
E.~R. Berlekamp, J.~H. Conway, R.~K. Guy, What is life?, in: Winning ways for
  your mathematical plays, Vol.~2, Academic Press, London, 1982.

\bibitem{liz10d}
J.~T. Lizier, M.~Prokopenko, A.~Y. Zomaya, Coherent information structure in
  complex computation, Theory in Biosciences 131 (2012) 193--203.
\newblock \href {http://dx.doi.org/10.1007/s12064-011-0145-9}
  {\path{doi:10.1007/s12064-011-0145-9}}.

\bibitem{Lindgren1990}
K.~Lindgren, M.~G. Nordahl, {Universal Computation in Simple One-Dimensional
  Cellular Automata}, Complex Systems 4~(3) (1990) 299--318.

\bibitem{Cook2008}
M.~Cook, A concrete view of rule 110 computation, in: T.~Neary, D.~Woods, A.~K.
  Seda, N.~Murphy (Eds.), {The Complexity of Simple Programs}, Vol.~1 of
  {EPTCS}, 2008, pp. 31--55.

\bibitem{weihrauch2000computable}
K.~Weihrauch, Computable Analysis: An Introduction {(Texts in Theoretical
  Computer Science. An EATCS Series)}, Springer, 2000.

\bibitem{Wolfram1984-b}
S.~Wolfram, Universality and complexity in cellular automata, Physica D 10
  (1984) 1--35.

\bibitem{Brady-Herken}
A.~H. Brady, {The Busy Beaver Game and the Meaning of Life}, in: R.~Herken
  (Ed.), A Half-century Survey on The Universal Turing Machine, Oxford
  University Press, Inc., New York, NY, USA, 1988, pp. 259--277.

\bibitem{wiki:octa}
{Wikipedia contributors},
  \href{http://www.conwaylife.com/w/index.php?title=OTCA_metapixel}{{OTCA}
  metapixel}, [Online; accessed 4-November-2018] (2018).
\newline\urlprefix\url{http://www.conwaylife.com/w/index.php?title=OTCA_metapixel}

\bibitem{pro09}
M.~Prokopenko, F.~Boschietti, A.~J. Ryan, An information-theoretic primer on
  complexity, self-organization, and emergence, Complexity 15~(1) (2009)
  11--28.

\bibitem{Kauffman2016}
S.~Kauffman, Humanity in a Creative Universe, Oxford University Press, New
  York, NY, USA, 2016.

\bibitem{casti94}
J.~L. Casti, Complexification: Explaining a Paradoxical World through the
  Science of Surprise, Harper Collins, New York, USA, 1994.

\bibitem{Markose2004b}
S.~M. Markose, {Novelty in complex adaptive systems {(CAS)} dynamics: a
  computational theory of actor innovation}, Physica A: Statistical Mechanics
  and its Applications 344~(1) (2004) 41--49.

\bibitem{prok2014}
M.~Prokopenko, Grand challenges for computational intelligence, Frontiers in
  Robotics and AI 1 (2014) 2.

\bibitem{Markose2017}
S.~M. Markose, Complex type 4 structure changing dynamics of digital agents:
  {Nash} equilibria of a game with arms race in innovations, Journal of
  Dynamics and Games 4~(3) (2017) 255--284.

\bibitem{binmore1987}
K.~Binmore, Modeling rational players: {Part I}, Economics and Philosophy 3~(2)
  (1987) 179--214.

\bibitem{harre2017}
M.~Harr{\'e}, Utility, revealed preferences theory, and strategic ambiguity in
  iterated games, Entropy 19~(5) (2017) 201.

\end{thebibliography}

\section*{Author contributions statement}

M.P. and F.B. conceived the idea of the paper and carried out initial analysis; 
M.P., M.H. and P.P. carried out analysis for section \ref{FS};
M.P., M.H., J.L. and F.B. carried out analysis for section \ref{TM};
M.P., M.H. and J.L. carried out analysis for section \ref{CA}; 
S.K. discussed the role of unprestatable functions vs undecidable dynamics. 
M.P and M.H. wrote the manuscript. 
All authors reviewed the manuscript.

\end{document}